\documentclass[journal,comsoc]{IEEEtran}

\IEEEoverridecommandlockouts

\usepackage[active]{srcltx} 
\usepackage{graphicx}
\usepackage{balance}
\usepackage{cite}
\usepackage{amssymb}
\usepackage[cmex10]{amsmath}
\usepackage{amsfonts}
\usepackage{amsthm}

\usepackage{eurosym}

\usepackage{booktabs}
\usepackage[bookmarks=false,draft]{hyperref} 

\usepackage{multirow}

\usepackage{algorithm}
\usepackage{algorithmicx}
\usepackage{algpseudocode}

\usepackage{subcaption}

\usepackage{braket}
\usepackage{tikz}
\usetikzlibrary{quantikz}
\usepackage{adjustbox}
\usepackage{easyReview}

\usepackage{rotating}

\usepackage{placeins}

\usepackage{dblfloatfix}

\begin{document}

\newcommand{\meterCom}{\scalebox{.5}{\begin{tikzcd} \meter{} \end{tikzcd}}}
\newcommand{\cwCom}{\scalebox{.5}{\begin{tikzcd} \pgfmatrixnextcell \cw \end{tikzcd}}}

\newcommand{\eqdef}{\stackrel{\triangle}{=}}


\newtheorem{theo}{Theorem}
\newtheorem{theor}{Theorem}
\newtheorem{cor}{Corollary}
\newtheorem{lem}{Lemma}
\newtheorem{prop}{Proposition}
\newtheorem{ins}{Insight}

\theoremstyle{remark}

\theoremstyle{definition}
\newtheorem{defin}{Definition}
\newtheorem{ass}{Assumption}
\newtheorem*{rem}{Remark}

\renewcommand{\qed}{$\blacksquare$}

\renewcommand{\algorithmiccomment}[1]{// #1}

\title{Quantum Switch for the Quantum Internet:\\ Noiseless Communications through Noisy Channels}

\author{Marcello~Caleffi,~\IEEEmembership{Senior~Member,~IEEE,}
	Angela~Sara~Cacciapuoti,~\IEEEmembership{Senior~Member,~IEEE}
	\thanks{M. Caleffi and A. S. Cacciapuoti are with the Department of Electrical Engineering and Information Technology (DIETI), University of Naples Federico II, Naples, 80125 Italy (e-mail: marcello.caleffi@unina.it; angelasara.cacciapuoti@unina.it).}
	\thanks{The authors are also with the Laboratorio Nazionale di Comunicazioni Multimediali, National Inter-University Consortium for Telecommunications (CNIT), Naples, 80126 Italy.}
}



\maketitle

\begin{abstract}
Counter-intuitively, quantum mechanics enables quantum particles to propagate simultaneously among multiple space-time trajectories. Hence, a quantum information carrier can travel through different communication channels in a quantum superposition of different orders, so that the relative time-order of the communication channels becomes indefinite. This is realized by utilizing a quantum device known as \textit{quantum switch}. In this paper, we investigate, from a communication-engineering perspective, the use of the quantum switch within the \textit{quantum teleportation process}, one of the key functionalities of the Quantum Internet.
Specifically, a theoretical analysis is conducted to quantify the performance gain that can be achieved by employing a quantum switch for the entanglement distribution process within the quantum teleportation with respect to the case of absence of quantum switch. This analysis reveals that, by utilizing the quantum switch, the quantum teleportation is heralded as a noiseless communication process with a probability that, remarkably and counter-intuitively, increases with the noise levels affecting the communication channels considered in the indefinite-order time combination. 
\end{abstract}



\section{Introduction}
\label{sec:1}

\begin{figure*}[h!]
	\centering
	\begin{minipage}[c]{.32\linewidth}
		\centering
		\includegraphics[width=.9\columnwidth]{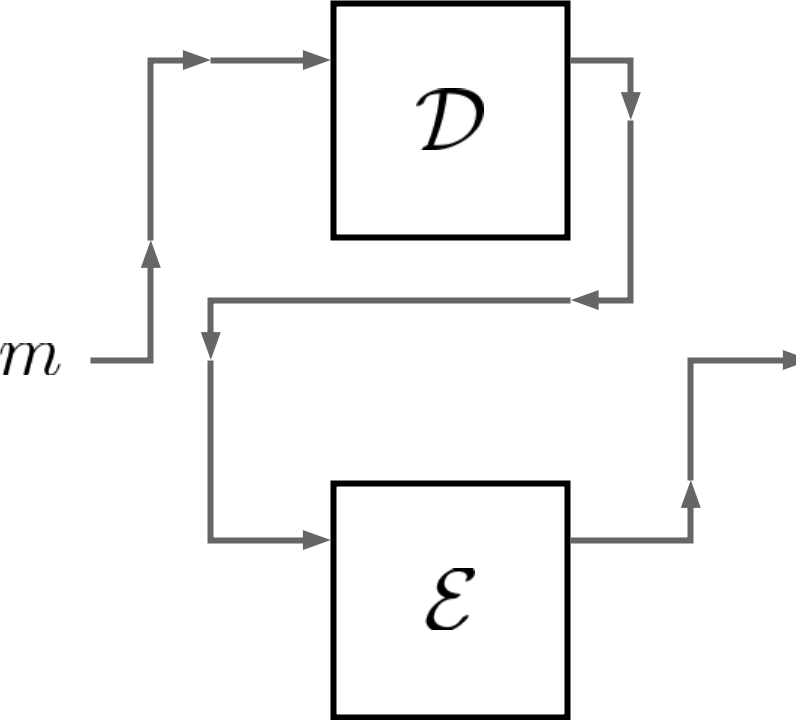}
		\subcaption{Message $m$ traverses channel $\mathcal{E}$ after channel $\mathcal{D}$, resulting in the transformation $\mathcal{E}(\mathcal{D}(m))$.}
		\label{Fig:1}
	\end{minipage}
	\hfill\hspace{3pt}
	\begin{minipage}[c]{.32\linewidth}
		\centering
		\includegraphics[width=.9\columnwidth]{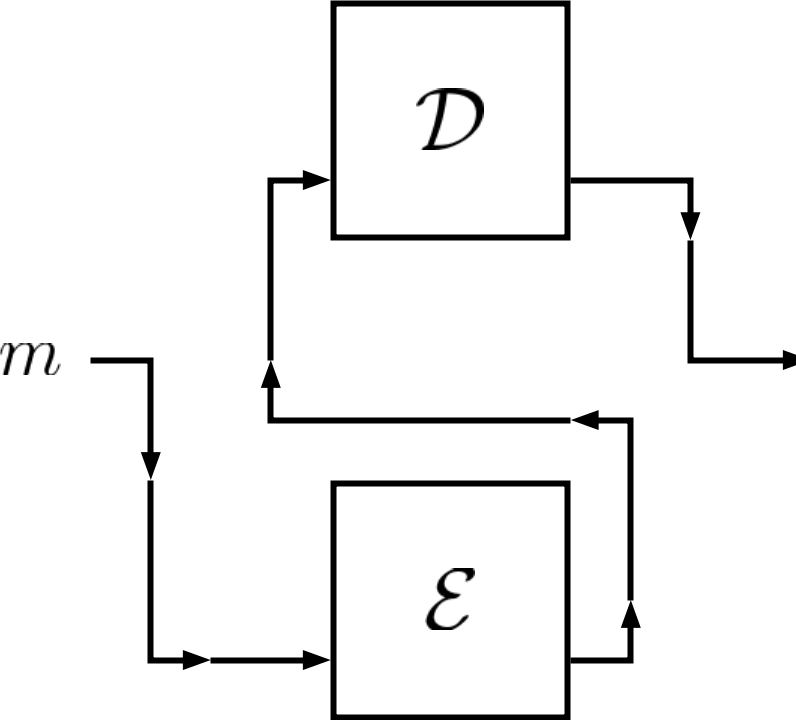}
		\subcaption{Message $m$ traverses channel $\mathcal{E}$ before channel $\mathcal{D}$, resulting in the transformation $\mathcal{D}(\mathcal{E}(m))$.}
		\label{Fig:2}
	\end{minipage}
	\hfill\hspace{3pt}
	\begin{minipage}[c]{.32\linewidth}
		\centering
		\includegraphics[width=.9\columnwidth]{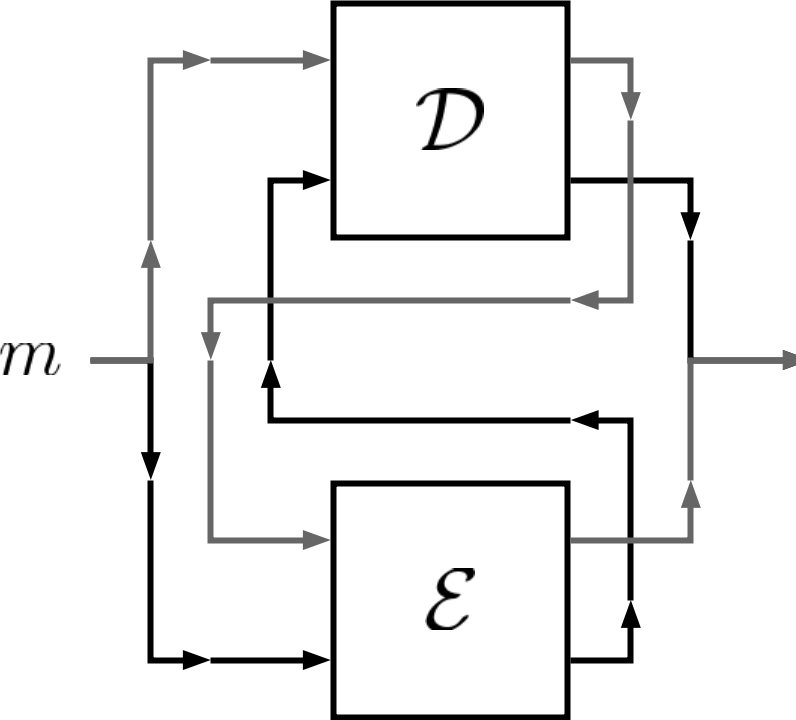}
		\subcaption{Message $m$ traverses the channels $\mathcal{D}$ and $\mathcal{E}$ in a superposition of the two alternative orders $\mathcal{D \rightarrow E}$ and $\mathcal{E \rightarrow D}$.}
		\label{Fig:3}
	\end{minipage}
	\caption{Pictorial representation of temporal trajectories: (a)-(b) a message traversing two channels in a well-defined temporal order; (c) a message traversing two channels in a superposition of different temporal orders.}
	\label{Fig:0}
\end{figure*}

Traditionally, the transmission of quantum information is assumed to flow along classical trajectories, i.e., trajectories that obey to the law of classical physics. Specifically, the quantum information carriers are usually assumed to travel along \textit{well-defined} trajectories in space-time \cite{SalEblChi-18}.

This assumption implies that, when the quantum message is sent through a sequence of communication channels, the order in which the channels are traversed is well-defined. As instance, with reference to Fig.~\ref{Fig:0}, when a message $m$ must go through two communication channels -- let us say channels $\mathcal{D}$ and $\mathcal{E}$ -- to reach the destination, either channel $\mathcal{E}$ is traversed after channel $\mathcal{D}$ as in Fig.~\ref{Fig:1} or vice versa as in Fig.~\ref{Fig:2}.

However, quantum particles can also propagate simultaneously among multiple space-time trajectories \cite{ChiKri-19}. This ability enables in principle the possibility for a quantum particle to experience a set of evolutions in a superposition of alternative orders\footnote{Indeed, a quantum particle can also experience a set of alternative evolutions by propagating simultaneously along multiple paths \cite{AbbWecHor-18}.}. In other words, quantum mechanics enables communication channels to be combined in time in \textit{a quantum superposition of different orders} as in Fig.~\ref{Fig:3}, with the relative order of the communication channels becoming indefinite.

This ``exotic'' communication scenario -- realized through a novel quantum device called \textit{quantum switch} \cite{ChiDarPer-13} -- arises when the temporal order of the communication channels is controlled by a quantum degree of freedom, represented by a \textit{control qubit}.

The utilization of a quantum switch provides significant advantages for a number of problems, ranging from quantum computation \cite{ColDarFac-12,ChiDarPer-13,AraCosBru-14} and quantum information processing \cite{Chi-12,WakSoeMur-19} through non-local games \cite{OreCosBru-12} to communication complexity \cite{FeiAraBru-15,GueFeiAra-16}. And multiple physical implementations of the quantum switch have been proposed and experimentally realized with photons \cite{ProMoqAra-15,RubRozFei-17,RubTozMas-19} -- with the control qubit represented by polarization or orbital angular momentum degrees of freedoms.

Even more interesting from a communication-engineering perspective, the quantum switch has been recently applied to quantum communications. Specifically, the quantum channel capacity\footnote{Indeed, a superposition of alternative orders provides advantages also in terms of classical channel capacity, as investigated in \cite{EblSalChi-18,GueRubBru-19,GosRomWhi-18,WeiNorZha-19}.} when the message traverses noisy channels in a superposition of alternative orders has been investigated theoretically \cite{SalEblChi-18,ChiBanSan-18,GueRubBru-19,ChiKri-19} and experimentally \cite{GosGiaKew-18,GuoHuHou-18}. And the results are remarkable \cite{SalEblChi-18}: \textit{a quantum superposition of two alternative orders of noisy channels can behave as a perfect quantum communication channel, even if no quantum information can be sent throughout either of the component channels individually}.

In this paper, inspired by these recent works, we investigate the use of the quantum switch within the \textit{quantum teleportation process}, one of the key functionalities of the Quantum Internet, as recently surveyed in \cite{CacCalVanHan-19}.

Specifically, quantum teleportation \cite{CalCacBia-18,CacCalTaf-18,CacCalVanHan-19} constitutes a priceless strategy for ``transmitting" qubits \cite{BenWie-92,BenBraCre-93}, without either the physical transfer of the particle storing the qubit or the violation of the quantum mechanics principles. To realize the marvels of the quantum teleportation two resources are needed. One resource is classic: two classical bits must be transmitted from the source to the destination. The other resource is quantum: a pair of maximally-entangled qubits must be generated and shared between the two parties. As a consequence, the entanglement generation/distribution process plays a crucial role within the Quantum Internet.

Unfortunately, the quantum entanglement is a very fragile resource, easily degraded by noise \cite{CacCal-19,CacCalVanHan-19}. And any entanglement degradation maps into a degradation of the teleported quantum information. Nevertheless, as shown through the paper, the deleterious effects of noisy communication channels on the entanglement distribution can be significantly reduced by exploiting the quantum superposition of two alternative time-orders provided by a quantum switch. 

In this context we conduct a theoretically analysis that quantifies the gain that can be achieved by employing a quantum switch for the entanglement distribution in the quantum teleportation with respect to the case of absence of quantum switch. More in details, we derive closed-form expressions that link the teleported qubit at Bob's side to the degradations experienced by the entangled pair during the distribution process. And stemming from these expressions, we evaluate the average fidelity achievable by utilizing the quantum switch.  The theoretical analysis reveals that the possibility of a quantum particle to experience a set of evolutions in a superposition of alternative orders is key to enhance the fidelity of the teleported qubit. Specifically, by utilizing the quantum switch, the quantum teleportation is heralded as a noiseless communication process with a probability that, remarkably and counter-intuitively, increases with the noise levels affecting the communication channels considered in the indefinite-order time combination.

The rest of the paper is organized as follows. In Sec.~\ref{sec:2} we provide some preliminaries about the quantum switch. Then in Sec.~\ref{sec:3} we first discuss the quantum teleportation process and the crucial role played by the entanglement generation and distribution process within the Quantum Internet, and then we introduce a practical communication system model for entanglement distribution through quantum switch. In Sec.~\ref{sec:4} we present some preliminaries on the entanglement distribution process realized through a quantum switch, whereas in Sec.~\ref{sec:5} we conduct the theoretical analysis of the quantum teleportation in presence of quantum switch. Finally in Sec.~\ref{sec:6} we conclude the paper by highlighting some challenges and open problems arising with the quantum switch.

\section{Quantum Switch: Preliminaries}
\label{sec:2}

\begin{table*}[bp]
	\caption{Quantum Gates}
	\label{Tab:1}
	\centering
	\begin{tabular}{|l|c|c|c|c|c|c|c|}
		\toprule
		\textbf{Gate} & 
			Identity & 
			 \begin{tabular}{@{}c@{}}
				X \\ (NOT)
			\end{tabular}
			& Y & Z & Hadamard &
			 \begin{tabular}{@{}c@{}}
				Controlled-NOT \\ (CNOT)
			\end{tabular}
			\\
		\midrule
		\textbf{Symbol} &
			\begin{tikzcd} & \gate{I} & \qw \end{tikzcd}	 &
			\begin{tikzcd} & \gate{X} & \qw \end{tikzcd} &
			\begin{tikzcd} & \gate{Y} & \qw \end{tikzcd} &
			\begin{tikzcd} & \gate{Z} & \qw \end{tikzcd} &
			\begin{tikzcd} & \gate{H} & \qw \end{tikzcd} &
			\begin{tikzcd} & \ctrl{1} & \qw \\& \targ{} & \qw \end{tikzcd}	\\
		\midrule
		\textbf{Matrix} &
			$I = \begin{bmatrix} 1 & 0 \\ 0 & 1 \end{bmatrix}$ &
			$X = \begin{bmatrix} 0 & 1 \\ 1 & 0 \end{bmatrix}$ &
			$Y = \begin{bmatrix} 0 & -i \\ i & 0 \end{bmatrix}$ &
			$Z = \begin{bmatrix} 1 & 0 \\ 0 & -1 \end{bmatrix}$ &
			$H = \displaystyle \frac{1}{\sqrt{2}} \begin{bmatrix} 1 & 1 \\ 1 & -1 \end{bmatrix}$ &
			$\begin{bmatrix} 1 & 0 & 0 & 0 \\ 0 & 1 & 0 & 0 \\ 0 & 0 & 0 & 1 \\ 0 & 0 & 1 & 0 \end{bmatrix}$ \\
		\bottomrule
	\end{tabular}
\end{table*}

As mentioned in Section~\ref{sec:1}, the quantum switch is a novel quantum device allowing a quantum particle to experience a set of evolutions in a superposition of alternative orders \cite{SalEblChi-18,EblSalChi-18}. In this ``exotic'' communication scenario, the relative order of the communication channels becomes indefinite, since the channel temporal order is governed by a quantum degree of freedom, which can be represented without any loss in generality by a qubit $\ket{\varphi_c}$, named \textit{control qubit}.

More in details, whenever the control qubit is initialized to one of the basis states, say $\ket{\varphi_c} = \ket{0}$, the quantum switch enables the message $m$ to experience the classical trajectory $\mathcal{D \rightarrow E}$ -- representing channel $\mathcal{E}$ being traversed after channel $\mathcal{D}$ -- as shown in Fig.~\ref{Fig:1}. Similarly, whenever the control qubit is initialized to the other basis state, say $\ket{\varphi_c} = \ket{1}$, the quantum switch enables the message $m$ to experience the alternative classical trajectory $\mathcal{E \rightarrow D}$ -- representing channel $\mathcal{E}$ being traversed before channel $\mathcal{D}$ -- as shown in Fig.~\ref{Fig:2}.

Conversely, whenever the control qubit is initialized to a superposition of the basis states, such as $\ket{\varphi_c} = \ket{+}$, the message $m$ experiences a quantum trajectory -- i.e., it experiences a superposition of the two alternative evolutions $\mathcal{D \rightarrow E}$ and $\mathcal{E \rightarrow D}$ -- as shown in Fig.~\ref{Fig:3}. 

Indeed, as an example of the quantum switch advantages, let us consider an arbitrary qubit $\ket{\varphi}$ traversing two noisy quantum channels $\mathcal{D}$ and $\mathcal{E}$, and let us assume channel $\mathcal{D}$ being the \textit{bit-flip channel} and channel $\mathcal{E}$ being the \textit{phase-flip channel}. The bit-flip channel $\mathcal{D}$ flips the state of a qubit from $\ket{0}$ to $\ket{1}$ (and vice versa) with probability $p$, leaving the qubit unaltered with probability $1-p$:
\begin{equation}
	\label{eq:2.1}
	\mathcal{D}(\varphi) = (1-p) \varphi + p X \varphi,
\end{equation}
where $X$ denotes the X-gate in Table~\ref{Tab:1}. The phase-flip channel $\mathcal{E}$ introduces -- with probability $q$ -- a relative phase-shift of $\pi$ between the complex amplitudes $\alpha$ and $\beta$ of the qubit $\ket{\varphi} = \alpha \ket{0} + \beta \ket{1}$, leaving the qubit unaltered with probability $1-q$:
\begin{equation}
	\label{eq:2.2}
	\mathcal{E}(\varphi) = (1-q) \varphi + q Z \varphi\add{,}
\end{equation}
where $Z$ denotes the Z-gate in Table~\ref{Tab:1}. Taken individually, the quantum capacity $\mathcal{Q}(\cdot)$ of each channel is \cite{Wil-13}:
\begin{equation}
	\label{eq:2.3}
	\begin{aligned}
		Q(\mathcal{D}) = 1 - h_b(p),\\
		Q(\mathcal{E}) = 1 - h_b(q),\\
	\end{aligned}
\end{equation}
respectively, with $h_b(x) \eqdef - x \log{x} - (1-x) \log{(1-x)}$ denoting the binary entropy.

When the two channels are traversed in a well-defined order, the overall quantum capacity is lower than the minimum of the individual capacities \cite{Wil-13} -- a result referred to as \textit{bottleneck capacity}. Hence, with reference to the classical well-defined trajectory $\mathcal{D} \rightarrow \mathcal{E}$, it results:
\begin{align}
	\label{eq:2.4}
	Q(\mathcal{D} \rightarrow \mathcal{E}) \leq & \min\{ Q(\mathcal{D}), Q(\mathcal{E}) \} = \nonumber \\
		&= 1 - \max\{ h_b(p), h_b(q) \} 
\end{align}
and the same result holds for the classical trajectory $\mathcal{E} \rightarrow \mathcal{D}$.

In particular, by considering the scenario where $p = q = \frac{1}{2}$ we have that no quantum information can be sent through any classical trajectory traversing the channels $\mathcal{D}$ and $\mathcal{E}$. Indeed, no quantum information can be sent either through any single instance of the channels.

Conversely and astounding, the quantum trajectory constituted by an even superposition of the two alternative evolutions $\mathcal{D \rightarrow E}$ and $\mathcal{E \rightarrow D}$ behave as an ideal channel with probability $pq = \frac{1}{4}$ \cite{SalEblChi-18}, violating so the bottleneck inequality in \eqref{eq:2.4}.

Hence, in a nutshell, \textit{a quantum superposition of two alternative orders of noisy channels can behave as a perfect quantum communication channel, even if no quantum information can be sent throughout either of the component channels individually} \cite{SalEblChi-18}.

\section{Quantum Switch for the Quantum Internet}
\label{sec:3}
Here we apply the quantum switch to the enabling functionality of the Quantum Internet: the entanglement generation and distribution process. Specifically, we first introduce in Sec.~\ref{sec:3.1} the quantum teleportation process and we highlight the key role played by the entanglement generation and distribution process within the Quantum Internet. Stemming from this, in Sec.~\ref{sec:3.2} we design a practical communication system model for entanglement distribution through quantum switch.

\subsection{Quantum Teleportation}
\label{sec:3.1}

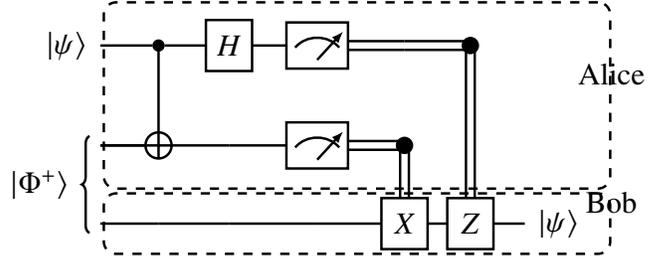
\begin{figure}[tp]
	\begin{adjustbox}{width=1\columnwidth}
		\begin{tikzcd}
			\\
			\lstick{$\ket{\psi}$}
				& \ctrl{1} \gategroup[wires=2,steps=7,style={dashed,rounded corners,inner xsep=10pt,inner ysep=3pt},label style={label position=right}]{Alice} & \gate{H} & \meter{} & \cw & \cwbend{2} & \\
			\lstick[wires=3]{$\ket{\Phi^+}$} & \targ & \qw & \qw & \meter{} & \cwbend{1} \\
			& \gategroup[wires=1,steps=7,style={dashed,rounded corners,inner xsep=10pt,inner ysep=3pt},label style={label position=right}]{Bob}
				\qw & \qw & \qw & \gate{X} & \gate{Z} & \qw \rstick{$\ket{\psi}$} &
		\end{tikzcd}
	\end{adjustbox}
	\caption{Quantum teleportation circuit. $\ket{\psi}$ denotes the qubit to be transmitted from Alice to Bob, and $\ket{\Phi^+}$ denotes the EPR pair generated and distributed so that one qubit is stored at Alice and another qubit is stored at Bob. The symbol \raisebox{.5 ex}{\protect\meterCom} denotes the measurement operation and the double-line \raisebox{.5 ex}{\protect\cwCom} represents the transmission of a classical bit from Alice to Bob.}
	\label{Fig:4}
\end{figure}

\textit{Quantum teleportation} \cite{CalCacBia-18,CacCalTaf-18,CacCalVanHan-19} constitutes a priceless strategy for ``transmitting" qubits \cite{BenWie-92,BenBraCre-93}, without either the physical transfer of the particle storing the qubit or the violation of the quantum mechanics principles.

To realize the marvels of the quantum teleportation two resources are needed. One resource is classic: two classical bits must be transmitted from the source -- say Alice -- to the destination -- say Bob. The other resource is quantum: a pair of maximally-entangled\footnote{In simple terms and oversimplifying, entanglement is a counter-intuitive form of correlation with no counterpart in the classical domain. By measuring individually any of the qubits forming the EPR pair, one obtains a random outcome. However, by comparing the results of the two independent measurements, one finds that they match, either directly or complementary. In particular, measuring one qubit of an EPR pair instantaneously changes the status of the second qubit, regardless of the distance dividing the two qubits \cite{CacCalVanHan-19}. For a more in-depth description of quantum entanglement and quantum teleportation, please refer to Sec.~II.D and Sec.~III in \cite{CacCalVanHan-19}.} qubits -- referred to as EPR pair in honor of Einstein, Podolsky, and Rosen's seminal work \cite{EinPodRos-35} -- must be generated and shared between Alice and Bob. 

Once the EPR pair is distributed between Alice and Bob, Alice performs a sequence of local operations on the two qubits at her side -- namely, the qubit to be teleported and one of the qubits forming the EPR pair -- as shown in Fig.~\ref{Fig:4}. Then, she transmits to Bob the output -- two classical bits -- of a joint measurement of the two qubits. Once Bob receives the two bits conveying Alice's measurement output, he can ``recover'' the original quantum information from the EPR qubit at his side with a sequence of local operations that depends on Alice's measurement, as depicted in Fig.~\ref{Fig:4}. It is worthwhile to note that, since the entanglement is destroyed during the teleportation process due to the measurement process, the teleportation of another qubit requires the generation and the distribution of a new EPR pair.

\begin{rem}
The entanglement generation/distribution process plays a key role within the Quantum Internet, since it is a fundamental pre-requisite for the transmission of quantum information through the quantum teleportation process.
\end{rem}

At this stage, a question arises: ``how an EPR pair can be generated and distributed between remote nodes?'' In a nutshell and by oversimplifying, the generation of quantum entanglement requires that two qubits interact each others, so that the state of each qubit cannot be described independently from the state of the other \cite{CacCalVanHan-19}. As an example, a popular scheme for entanglement generation involves carefully pointing a laser beam toward a non-linear crystal, so that two polarization-entangled photons emerge from the crystal \cite{KwiMatWei-95}.

Since Alice and Bob represents remote nodes, the entanglement generation occurring at one side must be complemented by the entanglement distribution functionality, which ``moves'' one of the entangled particles to the other side. To this matter, there is a broad consensus on the adoption of photons as entanglement carriers \cite{NorBla-17}. The rationale for this choice lays in the advantages provided by photons for entanglement distribution, such as weak interaction with the environment, easy control with standard optical components as well as high-speed low-loss transmission to remote nodes.

Despite the attractive features provided by photons as entanglement carriers, quantum entanglement is a very fragile resource and it is easily degraded by noise. More specifically, the effect of the noise is to transform the EPR pair into a non-maximally entangled pair, i.e., to \textit{degrade} the amount of entanglement shared between Alice and Bob. And any entanglement degradation introduces an unavoidable degradation\footnote{We refer the reader to \cite{CacCalVanHan-19} for a in-depth discussion about the different sources of imperfections affecting the quantum teleportation process.} of the quantum teleportation process, which becomes noisy.

\begin{rem}
The amount of entanglement degradation introduced during the entanglement generation/distribution process governs the imperfection of the teleportation process: the less the shared pair is entangled, the more the teleported qubit at Bob will differ from the original qubit at Alice.
\end{rem}

\subsection{Entanglement Distribution via Quantum Switch}
\label{sec:3.2}

\begin{figure}[tp]
	\centering
	\includegraphics[width=.9\columnwidth]{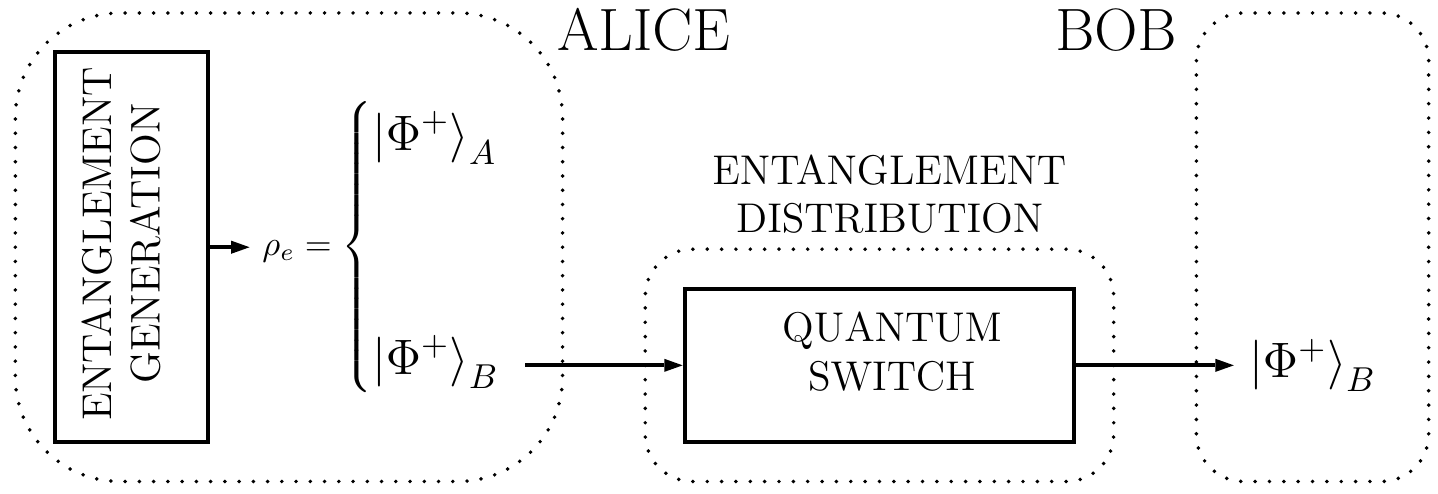}
	\caption{Entanglement distribution via quantum switch. The entanglement generation process is located at Alice, and $\rho_e = \ket{\Phi^+}\bra{\Phi^+}$ denotes the density matrix of the EPR pair $\ket{\Phi^+}$ generated at Alice. A quantum switch is employed to distribute the entanglement-pair member $\ket{\Phi^+}_B$ to Bob.}
	\label{Fig:5}
\end{figure}

With the discussions of Sec.~\ref{sec:3.1} in mind, here we aim at designing, from a communication-engineering perspective, a scheme able to exploit the quantum switch for entanglement distribution.

More in detail, we envision the scheme depicted in Fig.~\ref{Fig:5}. A pair of entangled particles is generated at Alice. Hence, one member of the EPR pair -- say $\ket{\Phi^+}_A$ -- is retained at Alice whereas the other member -- say $\ket{\Phi^+}_B$ -- is distributed to Bob through a quantum switch by using a photon as entanglement-carrier.

\begin{rem}
It is worthwhile to underline that the assumption of entanglement generation located \textit{at source} is not restrictive. Indeed, it constitutes one of most employed schemes for practical generation and distribution process as recently surveyed in \cite{CacCalVanHan-19}.
\end{rem}

Unfortunately, the quantum switch is an \textit{abstract function} rather than a well-defined physical device. Clearly, the naive implementation proposed in \cite{MukPat-19} does not fit with any practical communication system model, since it envisions a sequence of two teleportation processes sequentially applied in a superposition of time-orders. Furthermore, although a number of different physical implementations have been proposed in literature \cite{ProMoqAra-15,RubRozFei-17,GosGiaKew-18,GosRomWhi-18,GuoHuHou-18,WeiNorZha-19,RubTozMas-19}, these implementations aimed at confirming the theoretical results rather than at designing a communication system block. Indeed, within the mentioned implementations -- realized at a laboratory scale -- the communication links needed to interconnect the different components of a quantum switch were reasonably assumed ideal.

\begin{figure}[tp]
	\centering
		\begin{adjustbox}{width=1\columnwidth}
			\begin{tikzcd}
				\lstick[wires=2]{$\rho_e$} & [12pt]\lstick[label style={xshift=2}]{$\ket{\Phi^+}_A$} & \qw & \qw & \qw & \qw & \qw & \qw & \qw \rstick[wires=2]{$\rho_e^{\text{QS}}$}\\
				 & \lstick[label style={xshift=2}]{$\ket{\Phi^+}_B$} & \gategroup[wires=2,steps=5,style={dashed,rounded corners,inner xsep=10pt,inner ysep=3pt},label style={yshift=-3,label position=below}]{quantum switch} \gate[wires=2]{U} & \gate{\mathcal{D}} & \gate[wires=2]{SWAP} & \gate{\mathcal{D}} & \gate[wires=2]{U^{\dag}} & \qw & \qw \\
				\lstick{$\rho_c$} \qw & \qw & & \gate{\mathcal{E}} & & \gate{\mathcal{E}} & \qw & \qw & \qw \rstick{$\rho_c^{\text{out}}$}
			\end{tikzcd}
		\end{adjustbox}
	\caption{Circuit realization of a quantum switch for entanglement distribution. The entanglement-carrier $\ket{\Phi^+}_A$ is retained at Alice, whereas the entanglement-carrier $\ket{\Phi^+}_B$ is distributed at Bob through a quantum trajectory constituted by a superposition of alternative orders $\mathcal{D \rightarrow E}$ and $\mathcal{E \rightarrow D}$. The $U$ gate routes the entanglement-carrier $\ket{\Phi^+}_B$ either through channel $\mathcal{D}$ or $\mathcal{E}$, depending on the state of the control qubit $\varphi_c$. When the entanglement-carrier emerges from one channel, the $SWAP$ gate routes it through the other channel. Finally, the gate $U^\dag$ recombines the paths of the entanglement-carrier. $\rho_e^{\text{QS}}$ denotes the density matrix of the EPR pair distributed between Alice and Bob through the quantum switch.}
	\label{Fig:6}
\end{figure}
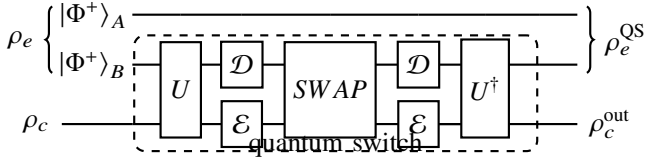

Conversely, we aim at modeling a practical communication system where any communication link -- regardless being an optical fiber link or a free-space optical link -- reasonably behaves as a noisy channel degrading the amount of entanglement eventually shared between Alice and Bob. Hence -- within the entanglement distribution framework -- we resort  to the circuit realization of the quantum switch given by the scheme in Fig.~\ref{Fig:6} and proposed in \cite{ChiKri-19}, which can be implemented with existing photonic technologies.

Specifically, in Fig.~\ref{Fig:6} the gate $U$ routes the entanglement-carrier $\ket{\Phi^+}_B$ through either the upper or the lower wire, depending on the state of the control qubit $\ket{\varphi_c}$. Regardless whether $\ket{\Phi^+}_B$ encountered channel $\mathcal{D}$ (upper wire) or channel $\mathcal{E}$ (lower wire), the $SWAP$ gate routes the entanglement-carrier through the other portion of the circuit -- hence realizing the trajectories $\mathcal{D} \rightarrow \mathcal{E}$ and $\mathcal{E} \rightarrow \mathcal{D}$, respectively. Eventually, regardless of the followed trajectory, gate $U^\dag$ routes the entanglement-carrier through the correct (upper) wire. Clearly, whenever the control qubit $\ket{\varphi_c}$ is in a superposition of the basis states, we have that the entanglement-carrier experiences the quantum trajectory corresponding to a superposition of the two alternative orders $\mathcal{D \rightarrow E}$ and $\mathcal{E \rightarrow D}$.

\begin{figure*}[tp]
	\centering
	\includegraphics[width=.8\linewidth]{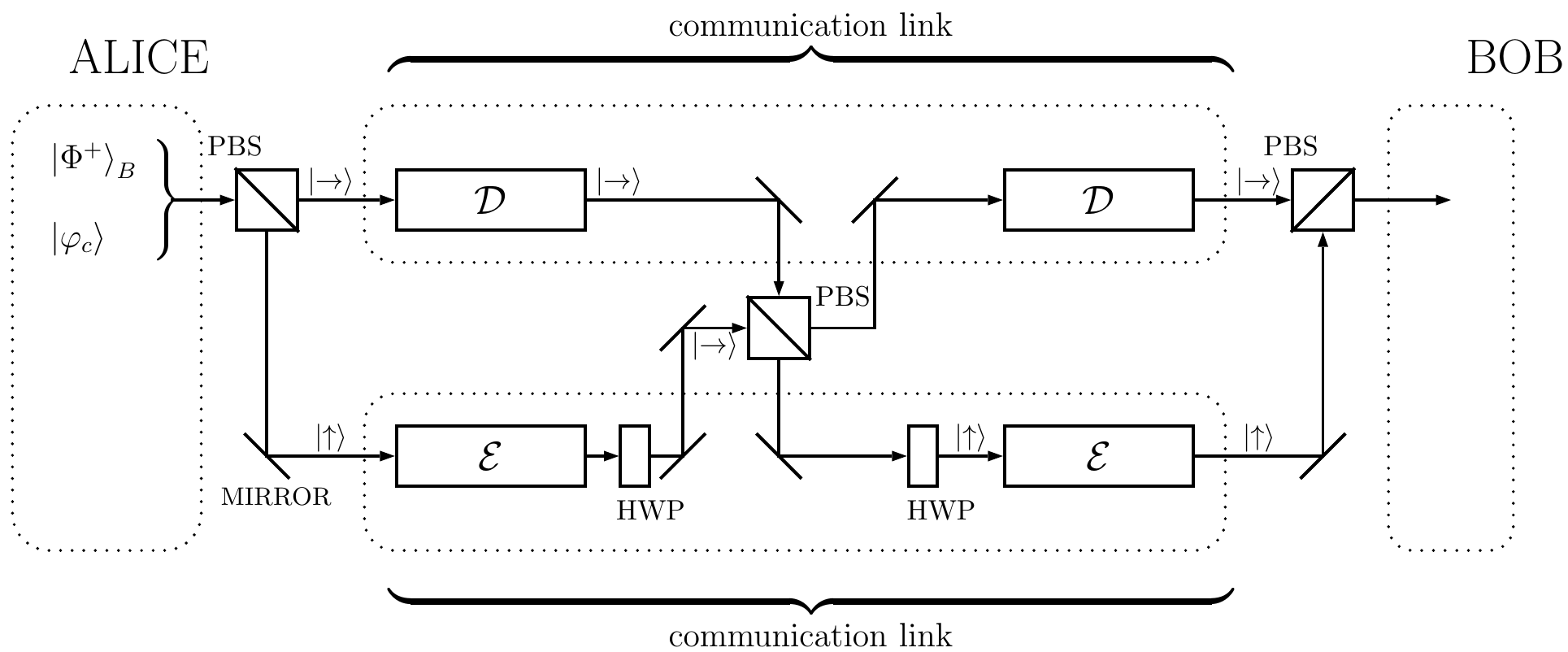}
	\caption{Sketch of a possible photonic implementation of a quantum switch for entanglement distribution. Two quantum communication links -- corresponding to the noisy quantum channels $\mathcal{D}$ and $\mathcal{E}$, respectively -- are available between Alice and Bob. The control qubit $\ket{\varphi_c}$ of the quantum switch is mapped into the horizontal/vertical photon's polarization $\ket{\rightarrow}$,$\ket{\uparrow}$, whereas the entangled-carrier $\ket{\Phi^+}_B$ is mapped into another photon's degree of freedom. The Polarization Beam Splitter (PBS) transmits a horizontally-polarized photon and reflects a vertically-polarized photon, whereas the Half-Wave Plate (HWP) realizes the polarization conversion between $\ket{\rightarrow}$ and $\ket{\uparrow}$.}
	\label{Fig:7}
\end{figure*}

From Fig.~\ref{Fig:6}, it becomes evident that a practical communication system model for entanglement distribution through quantum trajectories requires -- along with a $SWAP$ block -- two communication links. However, by mapping the control qubit $\ket{\varphi_c}$ and the entangled-carrier $\ket{\Phi^+}_B$ into different degrees of freedom of a single photon, it is possible to realize a quantum trajectory by transmitting a unique photon from Alice to Bob.

More in detail, in Fig.~\ref{Fig:7} we outline the sketch of a possible photonic implementation of a quantum switch for entanglement distribution -- with $\ket{\varphi_c}$ mapped into the photon's polarization $\ket{\rightarrow}$,$\ket{\uparrow}$ and $\ket{\Phi^+}_B$ mapped into another photon's degree of freedom. Whenever $\ket{\varphi_c}$ is initialized into a superposition of the basis states, two photons emerge from the first Polarization Beam Splitter (PBS): a horizontal-polarized photon and a vertical-polarized photon, which are sent to Bob through two different quantum communication links. The two photons, during their journey through the communication links, bump into a photonic $SWAP$ gate -- implemented with a PBS and a couple of Half-Wave Plates (HWPs) converting $\ket{\rightarrow}$ into $\ket{\uparrow}$ and vice versa -- which implements the superposition of alternative orders $\mathcal{D \rightarrow E}$ and $\mathcal{E \rightarrow D}$. Finally, the two photons emerging from the two paths are recombined at Bob with a third PBS.

\section{Modelling Entanglement Distribution via Quantum Switch}
\label{sec:4}

A quantum switch for a one-qubit system -- represented by the density matrix $\rho$ -- is described mathematically as a higher-order transformation \cite{SalEblChi-18} taking $\rho$ as input and returning as output $\mathcal{P}(\mathcal{D}, \mathcal{E}, \rho_c)(\rho)$, function of the two channel $\mathcal{D}$ and $\mathcal{E}$ along with the state $\rho_c=\ket{\varphi_c}\bra{\varphi_c}$ of the control qubit $\ket{\varphi_c}$:
\begin{align}
	\label{eq:4.1}
	\mathcal{P}(\mathcal{D}, \mathcal{E}, \rho_c)(\rho)= \sum_{i,j} W_{ij} (\rho \otimes \rho_c) W_{ij}^\dag.
\end{align}
In \eqref{eq:4.1}, $\{ W_{ij} \}$ denote the set of Kraus operators associated with the superposed channel trajectories, given by \cite{SalEblChi-18,ChiKri-19}:
\begin{align}
	\label{eq:4.2}
	W_{ij} = D_i E_j \otimes \ket{0}\bra{0} + E_j D_i \otimes \ket{1}\bra{1}.
\end{align}
with $\{D_i\}$ and $\{E_j\}$ denoting the Kraus operators associated with the channels $\mathcal{D}$ and $\mathcal{E}$, respectively.

Here, we extend this result to the entanglement distribution process. More in detail, by considering the circuital scheme depicted in Fig.~\ref{Fig:6} with photonic implementation given in Fig.~\ref{Fig:7}, we extend the use of the quantum switch to the case of a two-qubit system -- represented by the density matrix $\rho_e$ that is a $4 \times 4$ matrix. To this aim, we consider\footnote{As noted in \cite{SalEblChi-18}, this choice is not restrictive, since other types of depolarizing channels are unitarily equivalent to a bit flip and a phase flip channel. Hence the analysis can be easily extended by considering suitable pre-processing and post-processing operations.} the two noisy quantum channels introduced in Sec.~\ref{sec:2}: the \textit{bit flip channel} and the \textit{phase flip channel}, given in \eqref{eq:2.1} and \eqref{eq:2.2}, respectively.

By assuming without any loss of generality $\rho_e$ being the $4\times 4$ density matrix\footnote{We refer the reader to \cite{CacCalVanHan-19} for a concise introduction to the density matrix formalism, whereas a in-depth description can be found in \cite{NieChu-11}.} associated with the EPR pair $\ket{\Phi^+} = \left( \ket{00} + \ket{11} \right) / \sqrt{2}$:
\begin{equation}
	\label{eq:4.3}
	\rho_e\eqdef \ket{\Phi^+}\bra{\Phi^+} = \begin{bmatrix}
		\frac{1}{2} & 0 & 0 & \frac{1}{2} \\
		0 & 0 & 0 & 0 \\
		0 & 0 & 0 & 0 \\
		\frac{1}{2} & 0 & 0 & \frac{1}{2} 
	\end{bmatrix},
\end{equation}
we have the following results.

\begin{lem}
	\label{lem:4.1}
	The global quantum state $\mathcal{P}(\mathcal{D},\mathcal{E},\rho_c)(\rho_e)$ at the output of the quantum switch depicted in Fig.~\ref{Fig:6} is given by equation \eqref{eq:4.4} reported at the top of the next page, where $(i \oplus a)$ denotes the addition modulo-2 of $i$ and $a$.
	\begin{figure*}
		\begin{align}
			\label{eq:4.4}
			\mathcal{P}(\mathcal{D},\mathcal{E},\rho_c)(\rho_e)= &\left((1-p)(1-q) \rho_e + (1-p) q \left[\sum_{i} \left(\ket{i0}\bra{i0} - \ket{(i \oplus1) 1}\bra{(i\oplus 1)1}\right)\right] \rho_e \left[\sum_{i} \left(\ket{i0}\bra{i0} - \ket{(i \oplus1) 1}\bra{(i\oplus 1)1}\right)\right]^\dag+\right. \nonumber \\
				& \quad \left. +p(1-q)\left[\sum_{i,j} \ket{ij}\bra{i (j\oplus1)}\right] \rho_e \left[\sum_{i,j} \ket{ij}\bra{i (j\oplus1)}\right]^\dag\right) \otimes \ket{+}\bra{+} + \nonumber \\
				& + pq\left[\sum_{i,j} (-1)^{(j\oplus1)}\ket{ij}\bra{i (j\oplus1)}\right] \rho_e \left[\sum_{i,j} (-1)^{(j\oplus1)}\ket{ij}\bra{i (j\oplus1)}\right]^\dag \otimes \ket{-}\bra{-}
		\end{align}
	\end{figure*}
	\begin{IEEEproof}
		See Appendix~\ref{app:1}.
	\end{IEEEproof}
\end{lem}

\begin{rem}
From \eqref{eq:4.4}, it is possible to recognize that the effect of the quantum switch on the control qubit $\varphi_c$ with initial density matrix $\rho_c$ is to transform the control qubit into a mixed state of the two basis states $\ket{-}$,$\ket{+}$.
\end{rem}

By exploiting Lemma~\ref{lem:4.1}, we can derive the expression of the density matrix $\rho_e^{\text{QS}}$ of the EPR pair distributed between Alice and Bob at the output of the quantum switch.
\begin{cor}
	\label{cor:4.1}
	The density matrix $\rho_e^{\text{QS}}$ of the EPR pair distributed between Alice and Bob via a quantum switch is given by \eqref{eq:4.5} reported at the top of the next page.
	\begin{figure*}
		\begin{align}
			\label{eq:4.5}
			\rho_e^{\text{QS}} &=
			\begin{cases}
				\rho_e, &\text{with probability} \,\,\, pq,\\
				\begin{aligned}[l]
					& \dfrac{(1-p)(1-q) \rho_e + p(1-q)\left[\sum_{i,j} \ket{ij}\bra{i (j\oplus1)}\right] \rho_e \left[\sum_{i,j} \ket{ij}\bra{i (j\oplus1)}\right]^\dag}{1-pq}+\\
					& \quad + \dfrac{(1-p) q \left[\sum_{i} \left(\ket{i0}\bra{i0} - \ket{(i \oplus1) 1}\bra{(i\oplus 1)1}\right)\right] \rho_e \left[\sum_{i} \left(\ket{i0}\bra{i0} - \ket{(i \oplus1) 				1}\bra{(i\oplus 1)1}\right)\right]^\dag}{1-pq}
				\end{aligned} & \text{otherwise}\\
			\end{cases}
		\end{align}
	\end{figure*}
	\begin{IEEEproof}
		See Appendix~\ref{app:2}.
	\end{IEEEproof}
\end{cor}

\begin{rem}
From Corollary~\ref{cor:4.1}, we have two cases. With probability $pq$ -- heralded by a measurement of the control qubit corresponding to the state $\ket{-}$ -- the entanglement distribution is a noiseless process. In fact, Bob receives the particle $\ket{\Phi_B^+}$ of the EPR pair without any error, being $\rho^{\text{QS}}_e = \rho_e$ as detailed in Appendix~\ref{app:2}. As a consequence, by utilizing the quantum switch, the entanglement distribution process is a heralded noiseless communication process with probability $pq$. Differently, with probability $1-pq$ -- heralded by a measurement of the control qubit corresponding to the state $\ket{+}$ -- the entanglement distribution is a noisy process being Bob's particle $\ket{\Phi_B^+}$ degraded by the noisy channels. Nevertheless, as it will be shown in Proposition~\ref{prop:5.1}, also in this case the quantum switch provides a considerable gain -- in terms of degradation reduction -- with respect to the case of absence of quantum switch. 
\end{rem}

Before concluding this section, we give with Corollary~\ref{cor:4.2} another intermediate result: the expression of the density matrix $\rho_e^{\text{CT}}$ of the EPR pair distributed between Alice and Bob through the classical trajectory $\mathcal{D} \rightarrow \mathcal{E}$.

\begin{cor}
	\label{cor:4.2}
	The density matrix $\rho_e^{\text{CT}}$ of the EPR pair distributed between Alice and Bob through the classical trajectory $\mathcal{D} \rightarrow \mathcal{E}$ is given by \eqref{eq:4.6} reported at the top of the next page.
	\begin{figure*}
		\begin{align}
			\label{eq:4.6}
			\rho_e^{\text{CT}} = &
				(1-p)(1-q) \rho_e + p(1-q)\left[\sum_{i,j} \ket{ij}\bra{i (j\oplus1)}\right] \rho_e \left[\sum_{i,j} \ket{ij}\bra{i (j\oplus1)}\right]^\dag + \nonumber \\
				& + (1-p) q \left[\sum_{i} \left(\ket{i0}\bra{i0} - \ket{(i \oplus1) 1}\bra{(i\oplus 1)1}\right)\right] \rho_e \left[\sum_{i} \left(\ket{i0}\bra{i0} - \ket{(i \oplus1) 	1}\bra{(i\oplus 1)1}\right)\right]^\dag + \nonumber \\
				& + pq \left[ \sum_{i,j} (-1)^{j\oplus1}\ket{ij}\bra{i (j\oplus1)} \right] \rho_e \left[\sum_{i,j} (-1)^{j\oplus1}\ket{ij}\bra{i (j\oplus1)}\right]^\dag
		\end{align}
	\end{figure*}
	\begin{IEEEproof}
		See Appendix~\ref{app:2.bis}.
	\end{IEEEproof}
\end{cor}

\begin{rem}
Indeed, the expression $\rho_e^{\text{CT}}$ given in \eqref{eq:4.6} holds for both the classical trajectories $\mathcal{D} \rightarrow \mathcal{E}$ and $\mathcal{E} \rightarrow \mathcal{D}$.
\end{rem}

\section{Quantum Teleportation via Quantum Switch}
\label{sec:5}

Here, we evaluate the performance gain achievable by distributing the entanglement via a quantum switch within the quantum teleportation process.

To this aim, in the following we first collect some definitions. Then, we prove the preliminary result reported in Lemma~\ref{lem:5.1}, revealing the closed-form expression of the density matrix of the teleported qubit at Bob's side as a function of the density matrix of the EPR pair shared between Alice and Bob. Such a result is mandatory to understand and to quantify how the communication noise impairments on the EPR distribution process affect the teleported qubit. Finally, stemming from this, we prove the main result in Proposition~\ref{prop:5.1}.

Let $\rho_{\psi} \eqdef \ket{\psi}\bra{\psi}$ be the $2\times 2$ density matrix of the unknown pure quantum state $\ket{\psi}=\alpha \ket{0} + \beta \ket{1}=\cos\left(\frac{\theta}{2}\right) \ket{0}+e^{i\phi}\sin\left(\frac{\theta}{2}\right) \ket{1}$ that Alice wants to "transmit" to Bob via the quantum teleportation process introduced in Sec.~\ref{sec:3.1}. In spherical coordinates, $\rho_{\psi}$ is equivalent to:
\begin{align}
	\label{eq:5.1}
	\rho_{\psi}& = \begin{bmatrix}
			\cos^2\left(\frac{\theta}{2}\right) &\cos\left(\frac{\theta}{2}\right) e^{-i\phi}\sin\left(\frac{\theta}{2}\right) \\
			\cos\left(\frac{\theta}{2}\right) e^{i\phi}\sin\left(\frac{\theta}{2}\right)& \sin^2\left(\frac{\theta}{2}\right)
		\end{bmatrix} = \begin{bmatrix}
			\rho_{\psi}^{11} & \rho_{\psi}^{12} \\
 			\rho_{\psi}^{21} & \rho_{\psi}^{22} 
		\end{bmatrix}
\end{align}

To stress the generality of Lemma~\ref{lem:5.1}, it is convenient to introduce the notation $\tilde{\rho}_e$ to denote the density matrix of the actual EPR pair distributed between Alice and Bob. The rational of this choice is that Lemma~\ref{lem:5.1} holds regardless of specific noise affecting the entanglement generation and distribution process. As instance and according to this, whenever the entanglement generation/distribution process is perfect, it results $\tilde{\rho}_e = \rho_e$ given in \eqref{eq:4.3}. With this in mind we provide the following definitions.

\begin{defin}
	\label{def:5.1}
	Let us denote with $\left\{ \tilde{\rho}_{e_{ij}} \right\}_{i,j=1,2}$ the four sub-block matrices arising by partitioning the $4\times4$ density matrix $\tilde{\rho}_e$ of the actual EPR pair shared between Alice and Bob into $2 \times 2$ block-matrices, i.e.:
	\begin{equation}
		\label{eq:5.2}
		\tilde{\rho}_e = \begin{bmatrix}
			\tilde{\rho}_{e_{11}} & \tilde{\rho}_{e_{12}}\\
			\tilde{\rho}_{e_{21}} & \tilde{\rho}_{e_{22}} 
		\end{bmatrix}.
	\end{equation}
\end{defin}
 

\begin{defin}
	\label{def:5.2}
	$\mathbf{1}_{ij}^A$ denotes the indicator function of the teleportation measurement process at Alice, i.e.:
	\begin{equation}
		\mathbf{1}_{ij}^A=
			\begin{cases}
				1, &\text{if Alice measures state $\ket{ij}$}\\
				0, &\text{otherwise}.
			\end{cases}
	\end{equation}
\end{defin}

\begin{lem}
	\label{lem:5.1}
	The density matrix $\rho_t$ of the teleported qubit at Bob's side is equal to:
	\begin{align}
		\label{eq:5.3}
		\rho_t &=\mathbf{1}_{00}^A\left[2 \left( \rho_{\psi}^{11} \tilde{\rho}_{e_{11}}+ \rho_{\psi}^{12} \tilde{\rho}_{e_{12}} + \rho_{\psi}^{21} \tilde{\rho}_{e_{21}}+ \rho_{\psi}^{22} \tilde{\rho}_{e_{22}}\right)\right]+ \nonumber\\
		& +\mathbf{1}_{01}^A\left[2 X \left( \rho_{\psi}^{22} \tilde{\rho}_{e_{11}}+ \rho_{\psi}^{21} \tilde{\rho}_{e_{12}}+ \rho_{\psi}^{12} \tilde{\rho}_{e_{21}}+ \rho_{\psi}^{11} \tilde{\rho}_{e_{22}}\right)X^\dag\right] + \nonumber \\	
		&+\mathbf{1}_{10}^A\left[2 Z\left( \rho_{\psi}^{11} \tilde{\rho}_{e_{11}} - \rho_{\psi}^{12} \tilde{\rho}_{e_{12}}- \rho_{\psi}^{21} \tilde{\rho}_{e_{21}}+ \rho_{\psi}^{22} \tilde{\rho}_{e_{22}}\right)Z^\dag\right]
		+ \nonumber\\
		& +\mathbf{1}_{11}^A\left[2 Z X \left( \rho_{\psi}^{22} \tilde{\rho}_{e_{11}} - \rho_{\psi}^{21} \tilde{\rho}_{e_{12}} - \rho_{\psi}^{12} \tilde{\rho}_{e_{21}}+ \rho_{\psi}^{11} \tilde{\rho}_{e_{22}}\right)(ZX)^\dag\right],
	\end{align}
	where $\rho_{\psi}^{ij}$ is given in \eqref{eq:5.1}, $\mathbf{1}_{ij}^A$ is defined in Def.~\ref{def:5.2}, $\tilde{\rho}_e$ denotes the density matrix of the EPR pair distributed between Alice and Bob, and $\tilde{\rho}_{e_{ij}}$ is defined in Def.~\ref{def:5.1}.
	\begin{IEEEproof}
		See Appendix~\ref{app:3}.
	\end{IEEEproof}
\end{lem}

\begin{rem}
The closed-form expression \eqref{eq:5.3} derived within Lemma~\ref{lem:5.1} holds regardless of the particulars of the entanglement generation and distribution process, as long as $\tilde{\rho}_e$ denotes the density matrix of the actual EPR pair distributed between Alice and Bob. Specifically, \eqref{eq:5.3} holds for both quantum trajectories arising with a quantum switch as well as classical trajectories. Furthermore, \eqref{eq:5.3} holds regardless of the specific noise (if any) affecting the quantum channel used for entanglement distribution.
\end{rem}

Indeed, whenever the entanglement generation and distribution process is perfect, the sub-matrices $\{\tilde{\rho}_{e_{ij}}\}_{i,j = 1,2}$ are given by \eqref{eq:4.3}, i.e.:
\begin{equation}
	\label{eq:5.4}
	\begin{aligned}
		&\tilde{\rho}_{e_{11}}=\rho_{e_{11}}=\begin{bmatrix}
			\frac{1}{2} & 0 \\
			0 & 0 \end{bmatrix}, \quad 
		&\tilde{\rho}_{e_{12}}=\rho_{e_{12}}= \begin{bmatrix}
			0 & \frac{1}{2} \\
			0 & 0 \end{bmatrix},\\
		&\tilde{\rho}_{e_{21}}=\rho_{e_{21}}= \begin{bmatrix}
			0 &0 \\
			\frac{1}{2} & 0 \end{bmatrix}, \quad
		&\tilde{\rho}_{e_{22}}=\rho_{e_{22}}=\begin{bmatrix}
			0 &0 \\
			0 & \frac{1}{2} \end{bmatrix}.
	\end{aligned}
 \end{equation}
In this case, from \eqref{eq:5.3}, it is easy to recognize that the density matrix $\rho_t$ of the teleported qubit coincides with the density matrix $\rho_{\psi}$ of the unknown pure quantum state $\ket{\psi}$ for every possible outcome of the measurement. Conversely, whenever the entanglement generation and distribution process is imperfect, \eqref{eq:5.3} continues to hold but the sub-matrices $\{\tilde{\rho}_{e_{ij}}\}_{i,j = 1,2}$ deviate from their ideal expressions as a consequence of the noise.

\begin{figure*}[h!]
	\centering
	\begin{minipage}[c]{.49\linewidth}
		\centering
		\includegraphics[width=.9\columnwidth]{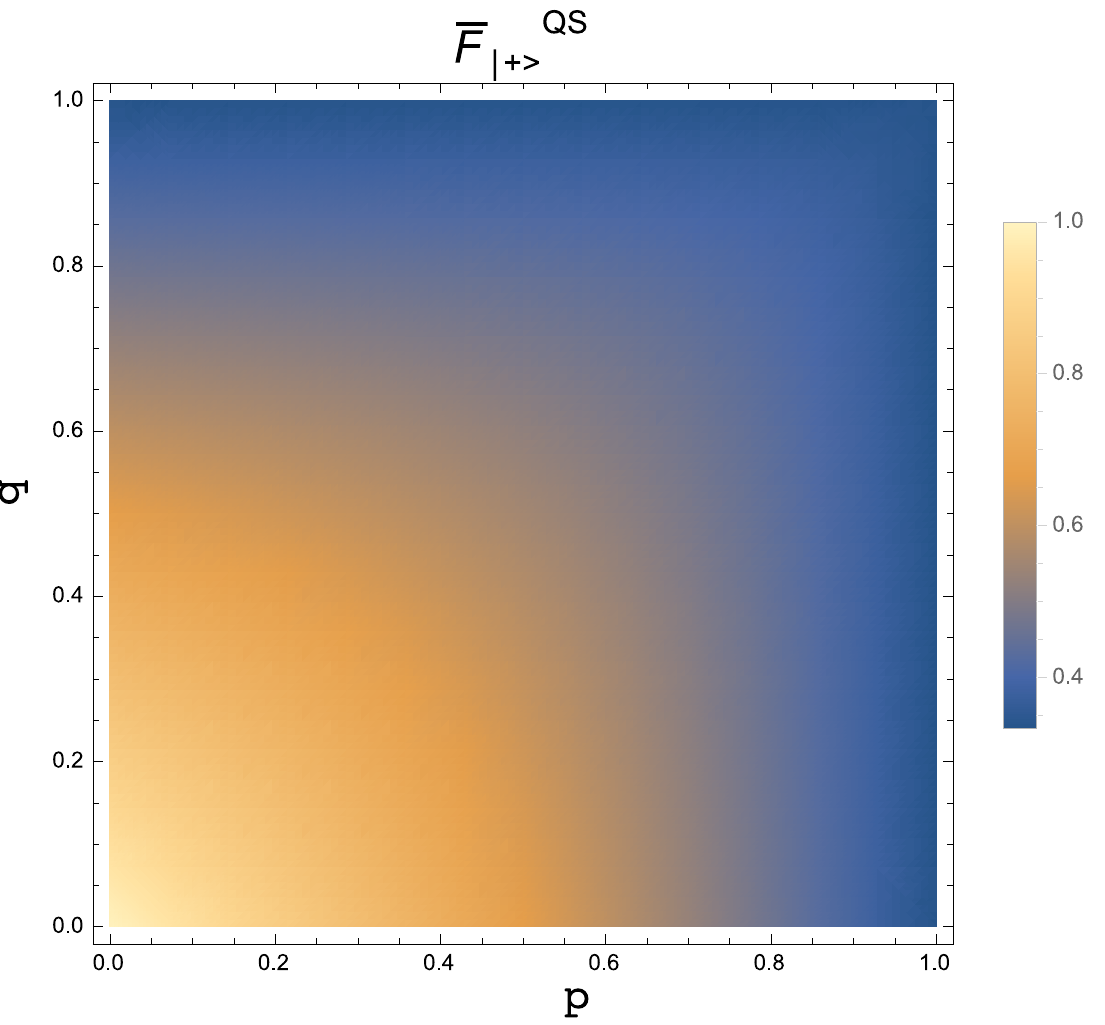}
		\subcaption{Average fidelity $\overline{F}_{\ket{+}}^{\text{QS}}$ given that the control qubit $\ket{\varphi_c}$ is measured into state $\ket{+}$.}
		\label{Fig:9}
	\end{minipage}
	\begin{minipage}[c]{.49\linewidth}
		\centering
		\includegraphics[width=.9\columnwidth]{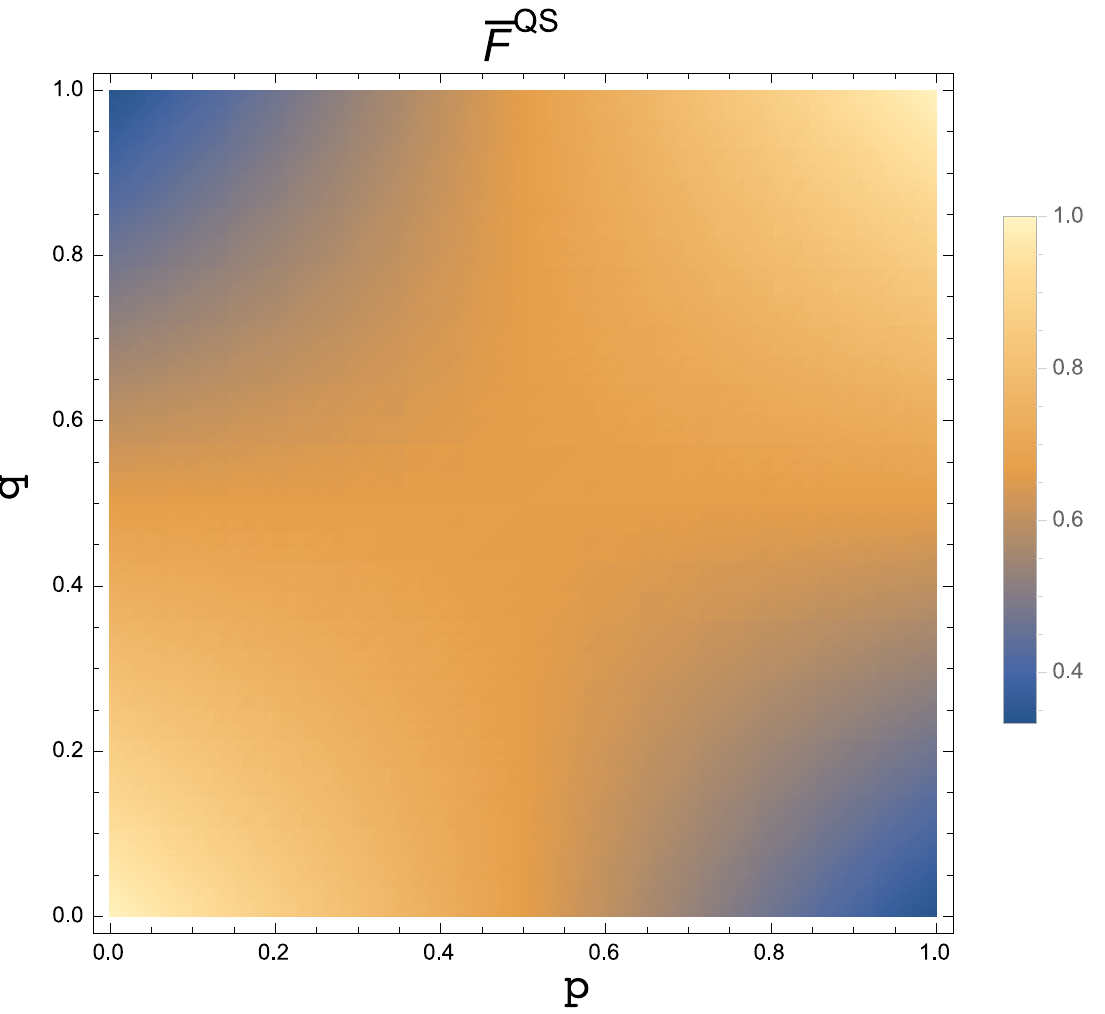}
		\subcaption{Average fidelity $\overline{F}^{\text{QS}} = pq \,\overline{F}_{\ket{-}}^{\text{QS}} + (1-pq) \overline{F}_{\ket{+}}^{\text{QS}}$.}
		\label{Fig:10}
	\end{minipage}
	\caption{Average fidelity of the teleported qubit when the EPR pair member $\ket{\Phi^+}_B$ is distributed at Bob's side via a quantum switch as a function of the error probabilities $p$ and $q$ of the two considered noisy channels $\mathcal{D}$ and $\mathcal{E}$ given in \eqref{eq:2.1} and \eqref{eq:2.2}.}
	\label{Fig:8}
\end{figure*}

In the following, stemming from the result derived in Lemma~\ref{lem:5.1}, we evaluate in Proposition~\ref{prop:5.1} and in the subsequent Corollary~\ref{cor:5.1} the performance gain -- in terms of reduction of the imperfections affecting the teleported qubit -- achievable through the superposition of casual orders via the quantum switch. For this, we resort to the fundamental figure of merit known as \textit{quantum fidelity} $F$. In a nutshell, the fidelity $F$ of an imperfect quantum state with density matrix $\rho$, with respect to a certain pure state $\ket{\psi}$, is a measure -- with values between 0 and 1 -- of the distinguishability of the two quantum states, and it is generally defined as $F = \bra{\psi} \rho \ket{\psi}$ \cite{Joz-94,VanMet-14}.

\begin{prop}
	\label{prop:5.1}
	The average fidelity $\overline{F}^{\text{QS}}$ of the teleported quantum state at Bob's side when the EPR pair is distributed via a quantum switch is given by:
	\begin{equation}
		\label{eq:5.5}
		\overline{F}^{\text{QS}}= \begin{cases}
				\overline{F}_{\ket{-}}^{\text{QS}} = 1, & \text{with probability} \,\,\, pq,\\
				\overline{F}_{\ket{+}}^{\text{QS}} =\frac{3-2p-2q+pq}{3(1-pq)}, & \text{otherwise}.
			\end{cases}
		\end{equation}
where $p$ and $q$ are the error probabilities of the two considered noisy channels $\mathcal{D}$ and $\mathcal{E}$ given in \eqref{eq:2.1} and \eqref{eq:2.2}, and $\overline{F}_{\ket{-}}^{\text{QS}}$ and $\overline{F}_{\ket{+}}^{\text{QS}}$ denote the average fidelity when the measurement of the control qubit correspond to the state $\ket{-}$ and $\ket{+}$, respectively.
	\begin{IEEEproof}
		See Appendix~\ref{app:4}
	\end{IEEEproof}
\end{prop}

\begin{rem}
	From \eqref{eq:5.5} it is easy to recognize that -- with probability $pq$ heralded by a measurement of the control qubit equal to $\ket{-}$ -- the quantum trajectory corresponding to a even superposition of the two alternative noisy evolutions $\mathcal{D \rightarrow E}$ and $\mathcal{E \rightarrow D}$ constitutes a noise-free channel. In fact, a fidelity equal to one -- which corresponds to the case of a teleported qubit at Bob identical to the original qubit to be teleported at Alice -- is obtained whenever the measurement of the control qubit returns state $\ket{-}$.
\end{rem}

\begin{rem}
\eqref{eq:5.5} can be equivalently written in a compact form as:
\begin{equation}
	\label{eq:5.6}
	\overline{F}^{\text{QS}}= pq \, \overline{F}_{\ket{-}}^{\text{QS}} + (1-pq) \overline{F}_{\ket{+}}^{\text{QS}} =\frac{3-2p-2q+4pq}{3}.
\end{equation} 
\end{rem}

\begin{figure*}[h!]
	\centering
	\begin{minipage}[c]{.49\linewidth}
		\centering
		\includegraphics[width=.9\columnwidth]{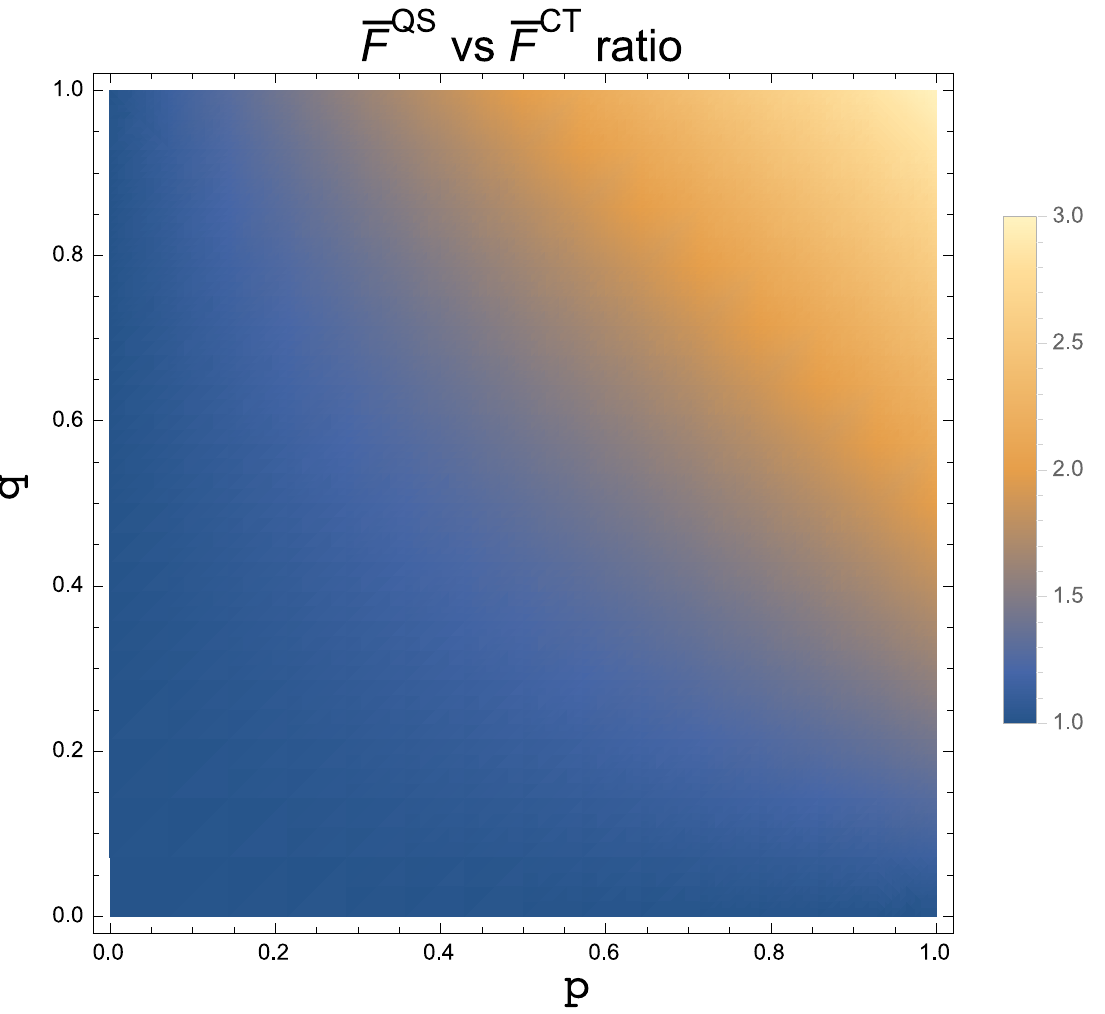}
		\subcaption{Ratio $\overline{F}^{\text{QS}} / \overline{F}^{\text{CT}}$ between the average fidelity of the teleported qubit when the EPR pair member $\ket{\Phi^+}_B$ is distributed to Bob: i) via a quantum switch, and and ii) through a classical trajectory.}
		\label{Fig:12}
	\end{minipage}
	\begin{minipage}[c]{.49\linewidth}
		\centering
		\includegraphics[width=1\columnwidth]{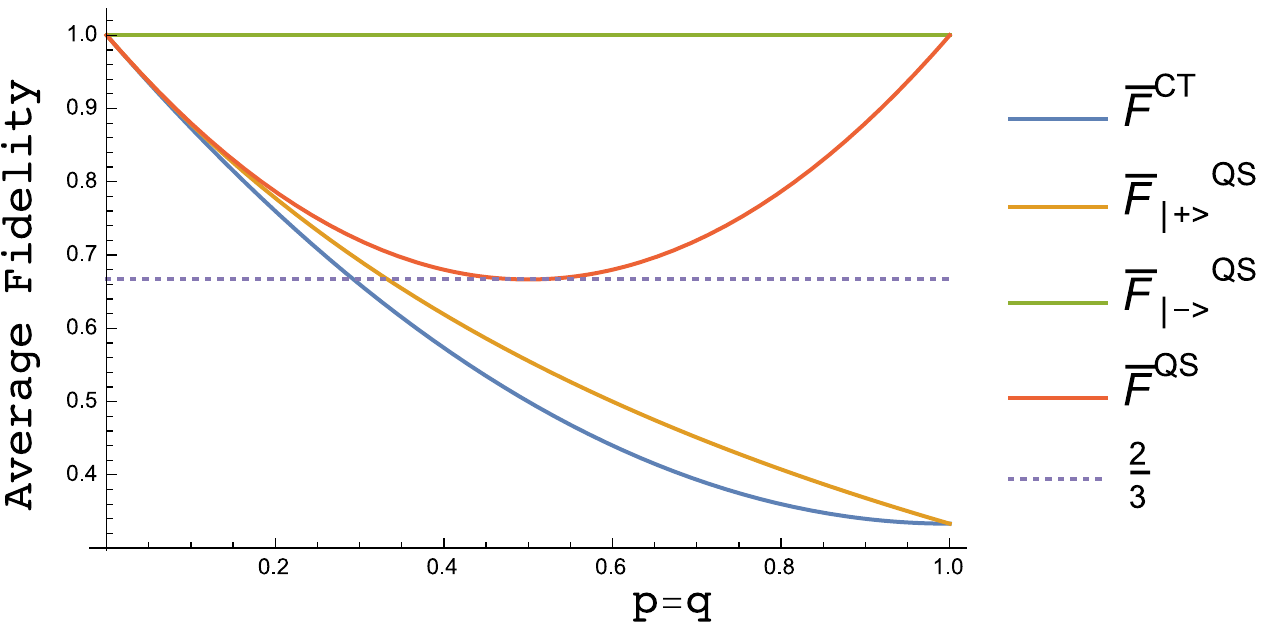}
		\subcaption{Average fidelity of the teleported qubit as a function of $p$ when $q=p$: i) $\overline{F}^{\text{CT}}$: average fidelity of the teleported qubit when the EPR pair member $\ket{\Phi^+}_B$ is distributed to Bob through a classical trajectory;
		ii) $\overline{F}_{\ket{+}}^{\text{QS}}$: average fidelity when the EPR pair member $\ket{\Phi^+}_B$ is distributed to Bob via a quantum switch, given that the control qubit $\ket{\varphi_c}$ is measured into state $\ket{+}$; iii) $\overline{F}_{\ket{-}}^{\text{QS}}$: average fidelity when the EPR pair member $\ket{\Phi^+}_B$ is distributed to Bob via a quantum switch, given that the control qubit $\ket{\varphi_c}$ is measured into state $\ket{-}$; iv) $\overline{F}^{\text{QS}}$: average fidelity when the EPR pair member $\ket{\Phi^+}_B$ is distributed to Bob via a quantum switch;}
		\label{Fig:13}
	\end{minipage}
	\caption{Performance comparison between the quantum and classical trajectories as a function of the error probabilities $p$ and $q$ of the two considered noisy channels $\mathcal{D}$ and $\mathcal{E}$ given in \eqref{eq:2.1} and \eqref{eq:2.2}.}
	\label{Fig:11}
\end{figure*}

Stemming from Proposition~\ref{prop:5.1}, in Fig.~\ref{Fig:8} we report the average fidelity achievable with a quantum switch, as a function of the error probabilities $p$ and $q$ of the two considered noisy channels -- i.e., the bit flip channel and the phase flip channel given in \eqref{eq:2.1} and \eqref{eq:2.2}, respectively. More in detail, in Fig.~\ref{Fig:9} we show the density plot of the average fidelity $\overline{F}_{\ket{+}}^{\text{QS}}$ obtained when the control qubit $\ket{\varphi_c}$ is measured into state $\ket{+}$ as a function of $p$ and $q$. As discussed within the remark following Corollary~\ref{cor:4.1}, whenever $\ket{\varphi_c}$ is measured into state $\ket{+}$, the noise on the quantum channels cause an unavoidable and irreversible degradation of the entanglement, which maps into a degradation of the teleported quantum information. This is evident in Fig.~\ref{Fig:9}: for any $p,q>0$ the average fidelity $\overline{F}_{\ket{+}}^{\text{QS}} < 1$, with values lower than $0.4$ for the highest values of the error probabilities $p$ and $q$. However, as it will be shown in the following with Fig.~\ref{Fig:11}, when a quantum switch is utilized for the entanglement distribution, the degradation of the teleported quantum state is always lower than the degradation introduced by the classical trajectory for any value of $p\neq 0$ and $q \neq 0$. As regards to the average fidelity $\overline{F}_{\ket{-}}^{\text{QS}}$ obtained whenever $\ket{\varphi_c}$ is measured into state $\ket{-}$, given that $\overline{F}_{\ket{-}}^{\text{QS}} = 1$ for any value of $p$ and $q$ a graphical plot is not necessary. Finally, in Fig.~\ref{Fig:10} we report the density plot of the average fidelity $\overline{F}^{\text{QS}} = pq \overline{F}_{\ket{-}} + (1-p) \overline{F}_{\ket{+}}^{\text{QS}}$ as a function of $p$ and $q$. It is worthwhile to note that the quantum switch guarantees an average fidelity exceeding the threshold $\frac{2}{3}$ -- that is the maximum fidelity achievable by distributing entanglement through classical channels \cite{MasPop-95} -- for most of the values spanned by $p$ and $q$. An exception arises whenever $p$ is close to zero and $q$ is close to $1$ (and vice versa). The rationale for this performance is that, in this case, the superposition of alternative orders collapses into a classical trajectory given that one of the two quantum channels behaves as an identical channel.

\begin{cor}
	\label{cor:5.1}
	The average fidelity $\overline{F}^{\text{QS}}$ of the teleported quantum state at Bob's side when the quantum switch is adopted is always greater than the average Fidelity $\overline{F}^{\text{CT}}$ of the teleported quantum state when a classical trajectory is adopted, for every $p,q \neq 0$:
	\begin{equation}
		\label{eq:5.7}
		\overline{F}^{\text{QS}} > \overline{F}^{\text{CT}}, 
	\end{equation} 
	where $\overline{F}^{\text{CT}} =\frac{3-2p-2q+2pq}{3}$.
	\begin{IEEEproof}
		See Appendix~\ref{app:5}
	\end{IEEEproof}
\end{cor}

Stemming from Corollary~\ref{cor:5.1}, in Fig.~\ref{Fig:11} we compare the average fidelities achievable with either a quantum switch or a classical trajectory, as a function of the error probabilities $p$ and $q$ of the bit flip and the phase flip channel, respectively. More in detail, in Fig.~\ref{Fig:12} we report the density plot of the ratio $\overline{F}^{\text{QS}} / \overline{F}^{\text{CT}}$ between $\overline{F}^{\text{QS}}$ -- the average fidelity of the teleported qubit when the EPR pair member $\ket{\Phi^+}_B$ is distributed to Bob via a quantum switch -- and $\overline{F}^{\text{CT}}$ -- the average fidelity of the teleported qubit when the EPR pair member $\ket{\Phi^+}_B$ is distributed to Bob through a classical trajectory. Indeed, Fig.~\ref{Fig:12} clearly shows the performance gain achievable by distributing entanglement via a quantum switch. To better visualize the performance gain in terms of fidelity, in Fig.~\ref{Fig:10} we plot the average fidelity as a function of $p$ when $q=p$. Remarkably, when $p=q=\frac{1}{2}$ -- i.e., when no quantum information can be sent through any classical trajectory traversing the channels $\mathcal{D}$ and $\mathcal{E}$ -- it results that $\overline{F}^{\text{QS}} = \frac{2}{3}$, whereas $\overline{F}^{\text{CT}} =\frac{1}{2}$. Furthermore, greater is the noise affecting the communications channels $\mathcal{D}$ and $\mathcal{E}$, higher is the performance gain in terms of fidelity provided by the quantum trajectory implemented via quantum switch. Differently, $\overline{F}^{\text{CT}}$ decreases by increasing $p$ and $q$. In the limit case of having $p=q \rightarrow 1$, $\overline{F}^{\text{CT}} \rightarrow \frac{1}{3}$ whereas $\overline{F}^{\text{QS}}\rightarrow 1$.

\begin{rem}
In a nutshell, distributing the entanglement through a quantum switch provides a significant performance gain -- in terms of fidelity of the teleported qubit at Bob's side -- for each level of the noise affecting the quantum communication channels. More remarkably, by retaining at Bob's side the entangled particles heralded by a $\ket{-}$-measurement of the control qubit $\ket{\varphi_c}$ and by discarding the particles heralded by a $\ket{+}$-measurement, \textit{the quantum switch realizes a noiseless entanglement distribution through noisy channels}.
\end{rem}

\section{Conclusions and Future Perspectives}
\label{sec:6} 
In this paper, we investigated the utilization of the quantum switch to face with the noise degradation introduced by the entanglement distribution within the quantum teleportation process.

The theoretical analysis revealed that exploiting the possibility for a quantum particle to experience a set of evolutions in a superposition of alternative orders is key to enhance the fidelity of the teleported qubit. Specifically, by utilizing the quantum switch, the teleportation is heralded as a noiseless communication process with a probability that, remarkably and counter-intuitively, increases with the noise levels affecting the communication channels considered in the indefinite-order time combination.

These preliminary results are encouraging. Nevertheless, a substantial amount of conceptual and experimental work has to be developed in order to tackle the challenges and open problems associated with the utilization of the quantum switch in the Quantum Internet. In the following, we outline some of these issues.

\subsubsection*{Quantum Switch vs Entanglement Distillation}
A well known technique to counteract the noise impairments affecting the entanglement generation/distribution process is the \textit{entanglement distillation} (or entanglement purification) \cite{Ben-96}. According to this technique, if the contamination of the entangled qubits is below a certain threshold, it is possible to purify multiple imperfectly entangled pairs into a single "almost-maximally entangled" pair, albeit at the price of additional processing. Hence, the entanglement purification exploits multiple transmissions of imperfect entangled pairs to obtain a single more entangled pair \cite{CacCalVanHan-19}.
By comparing entanglement purification and quantum switch from a communication network perspective, we can argue that the communication delay induced by the former seems to be higher that the one induced by the latter. However, further research is needed to quantify this possible delay-advantage. Finally, it is worthwhile to note that the benefits provided by the quantum switch and the entanglement purification can be mutually combined -- with the quantum switch enhancing the fidelity of each imperfect entangled pair, hence reducing the number of imperfect pairs required at the destination to distill a maximally entangled pair -- rather than constituting mutually-exclusive alternatives.

\subsubsection*{Network Design Issues}
The possible photonic implementation of a quantum switch sketched in Sec.~\ref{sec:3.2} requires the availability of two communication links between Alice and Bob, interconnected through a swapping device. Although multiple physical implementations of the quantum switch have been proposed in literature \cite{ProMoqAra-15,RubRozFei-17,RubTozMas-19,GosGiaKew-18,GuoHuHou-18} as discussed in Sec.~\ref{sec:1}, further research is needed to face with the challenges arising with the quantum network design. As instance, when the quantum communication links are implemented through optical fiber links, the interconnection through the quantum swap requires a spatial proximity between the fibers, which in turns poses additional constraints on the network topology.

\subsubsection*{Channel Noise}
The assumption of channel $\mathcal{D}$ being the \textit{bit-flip channel} and channel $\mathcal{E}$ being the \textit{phase-flip channel} is not restrictive, since other types of depolarizing channels are unitarily equivalent to a bit flip and a phase flip channel. Hence the analysis can be easily extended by considering suitable pre-processing and post-processing operations, as noted in \cite{SalEblChi-18}. Nevertheless, further research is needed to quantify the performance gain achievable when both the entangled qubits are distributed via quantum switches through noisy channels. 

Finally, the question whether the quantum switch can be integrated within the framework of quantum error correction techniques \cite{Han-QERC-18} is an open and interesting problem.

\appendices

\section{Proof of Lemma~\ref{lem:4.1}}
\label{app:1}

According to the entanglement distribution with circuital scheme depicted in Fig.~\ref{Fig:6} and photonic implementation given in Fig.~\ref{Fig:7}, the entanglement-pair member $\ket{\Phi^+}_A$,is already at Alice's side, thus it does not need to go throughout any communication channel. Differently, the second qubit of the EPR pair $\ket{\Phi^+}_B$ needs to be distributed to Bob.

By distributing $\ket{\Phi^+}_B$ through a quantum switch, the state of the global system constituted by the entangled pair $\rho_e$, the control qubit $\rho_c=\ket{+}\bra{+}$ and the communication channels $\mathcal{D}$ and $\mathcal{E}$ can be described through the Kraus operators $W_{ij}$ given by:
\begin{align}
	\label{eq:app.1.1}
	W_{ij} = &\left(I \otimes D_i\right) \left(I \otimes E_j\right) \otimes \ket{0}\bra{0} + \nonumber \\
		&\left(I \otimes E_j\right) \left(I \otimes D_i\right) \otimes \ket{1}\bra{1},
\end{align}
being the first qubit of the entangled pair (virtually) traveling throughout an ideal channel represented by the unitary transformation $I$ given in Table~\ref{Tab:1}.
By exploiting the tensor product properties, such as $A\otimes C + B \otimes C=(A+B) \otimes C$ and $(A_1\otimes B_1) (A_2 \otimes B_2) =A_1A_2\otimes B_1B_2$, \eqref{eq:app.1.1} can be rewritten equivalently as:
\begin{align}
	\label{eq:app.1.2}
	W_{ij} = I \otimes \left( D_i E_j \otimes \ket{0}\bra{0} + E_j D_i \otimes \ket{1}\bra{1}\right).
\end{align}
Since $\mathcal{D}$ and $\mathcal{E}$ denotes the bit flip channel and the phase flip channel, respectively, their Kraus operators are given by \cite{NieChu-11}:
\begin{equation}
	\label{eq:app.1.3}
	\begin{split}
		&D_1= (1-p) I, \quad D_2= p X \\
		&E_1= (1-q) I, \quad E_2= p Z. 
	\end{split}
\end{equation}
By substituting \eqref{eq:app.1.2} and \eqref{eq:app.1.3} in \eqref{eq:4.1}, and by exploiting again the tensor product properties, after some algebraic manipulations it results:
\begin{align}
	\label{eq:app.1.4}
	\mathcal{P}(\mathcal{D},\mathcal{E},\rho_c)(\rho_e)&=(1-p)(1-q)(\rho_e \otimes \ket{+}\bra{+}) +\\ &\nonumber 
		+(1-p)q (I \otimes Z) \rho_e (I \otimes Z) \otimes \ket{+}\bra{+} + \\&\nonumber 
		+ p(1-q) (I \otimes X) \rho_e (I \otimes X) \otimes \ket{+}\bra{+} +\\&\nonumber 
	+ pq (I \otimes XZ) \rho_e (I \otimes XZ)^\dag \otimes (Z\ket{+}\bra{+}Z )
\end{align}
From \eqref{eq:app.1.4}, the proof follows by recognizing that $Z\ket{+}\bra{+}Z=\ket{-}\bra{-}$ and that:
\begin{equation}
	\label{eq:app.1.5}
	\begin{split}
		& I \otimes Z = \sum_{i} \left(\ket{i0}\bra{i0} - \ket{(i \oplus1) 1}\bra{(i\oplus 1)1}\right) \\
		& I \otimes X = \sum_{i,j} \ket{ij}\bra{i (j\oplus1)}\\ 
		& I \otimes XZ = \sum_{i,j} (-1)^{j\oplus1}\ket{ij}\bra{i (j\oplus1)}
	\end{split}
\end{equation}

\section{Proof of Corollary~\ref{cor:4.1}}
\label{app:2}

From \eqref{eq:4.4} of Lemma~\ref{lem:4.1}, it results that the global state $\mathcal{P}(\mathcal{D},\mathcal{E},\rho_c)(\rho_e)$ at the output of the quantum switch is a mixture of pure states $\{\ket{+},\ket{-}\}$ of the control qubit. As a consequence -- by measuring the control qubit in the Hadamard basis -- whenever the measurement outcome is equal to $\ket{-}$, the global state collapses into the state $\rho_{e,\ket{-}}^{QS}$ reported in equation \eqref{eq:app.2.1} at the top of the next page. In this case -- happening with probability $pq$ -- Bob receives the particle $\ket{\Phi^+}_B$ of the EPR pair without any error. In fact, $\rho_e$ can be recovered perfectly from $\rho_{e,\ket{-}}^{\text{QS}}$ by simply applying on $\rho_{e,\ket{-}}^{\text{QS}}$ the unitary corrective operation $(I \otimes XZ )$, defined in \eqref{eq:app.1.5}.

\begin{figure*}
	\begin{align}
	\label{eq:app.2.1}
	\rho_{e,\ket{-}}^{QS} = \left[\sum_{i,j} (-1)^{(j\oplus1)}\ket{ij}\bra{i (j\oplus1)}\right] \rho_e \left[\sum_{i,j} (-1)^{(j\oplus1)}\ket{ij}\bra{i (j\oplus1)}\right]^\dag
\end{align}
\end{figure*}

Differently, when the measurement outcome of the control qubit is the one corresponding to the state $\ket{+}$, the global state collapses into the state $\rho_{e,\ket{+}}^{QS}$ reported in \eqref{eq:app.2.2} at the top of the next page. In this case -- happening with probability $(1-pq)$ -- Bob cannot receives the particle $\ket{\Phi_B^+}$ without error. Nevertheless, also in this case as it will be shown in Proposition~\ref{prop:5.1}, a considerable gain is assured with respect to the standard channel composition arising with classical trajectories. 

\begin{figure*}
\begin{align}
	\label{eq:app.2.2}
	\rho_{e,\ket{+}}^{QS} = & \frac{\displaystyle (1-p)(1-q) \rho_e + p(1-q)\left[\sum_{i,j} \ket{ij}\bra{i (j\oplus1)}\right] \rho_e \left[\sum_{i,j} \ket{ij}\bra{i (j\oplus1)}\right]^\dag}{1-pq} + \nonumber \\
		& + \frac{\displaystyle (1-p) q \left[\sum_{i} \left(\ket{i0}\bra{i0} - \ket{(i \oplus1) 1}\bra{(i\oplus 1)1}\right)\right] \rho_e \left[\sum_{i} \left(\ket{i0}\bra{i0} - \ket{(i \oplus1) 1}\bra{(i\oplus 1)1}\right)\right]^\dag}{1-pq}
\end{align}
\end{figure*}

\section{Proof of Corollary~\ref{cor:4.2}}
\label{app:2.bis}
When the bit-flip and phase-flip channels are traversed in a well defined order - let us say $\mathcal{D} \rightarrow \mathcal{E}$ - the density matrix of the entangled pair $\rho_e^{\text{CT}}$ at Bob's side is given by:
\begin{align}
	\label{eq:app.2.bis.1}
	\rho_e^{\text{CT}} & =\mathcal{E}\left[\mathcal{D}\left(\rho_e\right)\right]=\mathcal{E}\left[\sum_{i=1,2} D_i \rho_e D_i^\dag\right]=\\&\nonumber
= \sum_{j=1,2} E_j \left[\sum_{i=1,2} D_i \rho_e D_i^\dag\right] E_j^\dag.
\end{align}
By substituting \eqref{eq:app.1.3} in \eqref{eq:app.2.bis.1} and by accounting for \eqref{eq:app.1.5}, the proof follows after some algebraic manipulations.

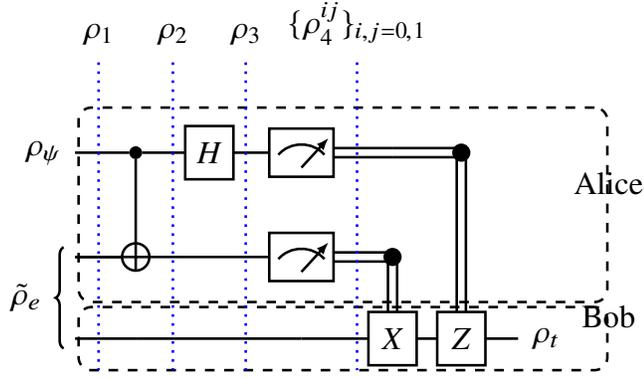
\begin{figure}[tp]
	\begin{adjustbox}{width=1\columnwidth}
		\begin{tikzcd}
			\\
			\lstick{$\rho_{\psi}$}
				\slice[style={blue,dotted},label style={inner xsep=-25pt,inner ysep=5}]{$\rho_1$}
				&\ctrl{1}
				\slice[style={blue,dotted},label style={inner xsep=-25pt,inner ysep=5}]{$\rho_2$}
				\gategroup[wires=2,steps=7,style={dashed,rounded corners,inner xsep=10pt,inner ysep=3pt},label style={label position=right}]{Alice}
				&\gate{H}
				\slice[style={blue,dotted},label style={inner xsep=-25pt,inner ysep=5}]{$\rho_3$}
				&\meter{}
				\slice[style={blue,dotted},label style={inner xsep=-25pt,inner ysep=5}]{$\{\rho_4^{ij}\}_{i,j=0,1}$}
				& \cw & \cwbend{2} & \\
			\lstick[wires=3]{$\tilde{\rho}_{e}$} & \targ & \qw & \qw & \meter{} & \cwbend{1} \\
			&
				\gategroup[wires=1,steps=7,style={dashed,rounded corners,inner xsep=10pt,inner ysep=3pt},label style={label position=right}]{Bob}
				\qw & \qw & \qw & \gate{X} & \gate{Z} & \qw \rstick{$\rho_t$} & &
		\end{tikzcd}
	\end{adjustbox}
	\caption{Pictorial Representation of the Quantum Teleportation Process in terms of density matrices.
	}
	\label{Fig:A}
\end{figure}

\section{Proof of Lemma~\ref{lem:5.1}}
\label{app:3}

To prove the lemma, let us consider Fig.~\ref{Fig:A} in which we depicted schematically the quantum teleportation process. The initial global state $\rho_1 \in \mathbb{C}^{8 \times 8}= \rho_{\psi} \otimes \tilde{\rho}_e$ is an $8 \times 8$ matrix given by :
\begin{equation}
	\label{app:3.1}
	\rho_1 =\rho_{\psi} \otimes \tilde{\rho}_e=
		\begin{bmatrix}
			\rho_{\psi}^{11} \tilde{\rho}_e & \rho_{\psi}^{12} \tilde{\rho}_e \\
			\rho_{\psi}^{21} \tilde{\rho}_e & \rho_{\psi}^{22} \tilde{\rho}_e
	\end{bmatrix}.
\end{equation}
As indicated in the main text, we denoted with $\tilde{\rho}_e$ the entanglement density matrix, since we do not formulate any assumption on the scheme employed for the entanglement generation and distribution process as well as for the noise affecting the process. Specifically, $\tilde{\rho}_e$ can be either given by \eqref{eq:4.3} in absence of noise or can be in some way affected by the noise.

The teleportation process starts with Alice applying the CNOT-gate of Table~\ref{Tab:1} to the pair of qubits at her side. In terms of density matrix, this is equivalent to consider an unitary operator $U=C_{\text{NOT}} \otimes I_{2\times2}$ acting on the global state $\rho_1$, so that the Bob's qubit is left unchanged:
\begin{equation}
	\label{app:3.2}
	\rho_2 = U \rho_1 U^\dag=(C_{\text{NOT}} \otimes I_{2\times2})\rho_1(C_{\text{NOT}} \otimes I_{2\times2}).
\end{equation}
By accounting for the expression of the CNOT gate and by substituting \eqref{app:3.1} in \eqref{app:3.2}, after some algebraic manipulations one obtains:
\begin{align}
	\label{app:3.3}
	\rho_2 & = \begin{bmatrix}
		\rho_{\psi}^{11} \tilde{\rho}_e & \rho_{\psi}^{12} \tilde{\rho}_e \chi \\
		\rho_{\psi}^{21} \chi \tilde{\rho}_e & \rho_{\psi}^{22}\chi \tilde{\rho}_e \chi
	\end{bmatrix},
\end{align}
with $\chi \in \mathbb{R}^{4\times4}$ equal to:
\begin{equation}
	\label{app:3.4}
	\chi \eqdef \begin{bmatrix}0_{2\times2} & I_{2\times2} \\ I_{2\times2} &0_{2\times2}\end{bmatrix}=\chi^\dag.
\end{equation}

Then, as shown in Fig.~\ref{Fig:A}, Alice applies the H-gate of Table~\ref{Tab:1} to the state to be teleported. Hence the global state $\rho_3$ after the H gate is:
\begin{equation}
	\label{app:3.5}
	\rho_3 = (H \otimes I_{2\times2} \otimes I_{2\times2} ) \rho_2 (H \otimes I_{2\times2} \otimes I_{2\times2} )^\dag.
\end{equation}
By accounting for the expression of the H gate, it results:
\begin{equation}
	\label{app:3.6}
	H \otimes I_{2\times2} \otimes I_{2\times2} = \frac{1}{\sqrt{2}}\begin{bmatrix}I_{4\times4} & I_{4\times4}\\I_{4\times4} & -I_{4\times4}\end{bmatrix}.
\end{equation}
By substituting \eqref{app:3.6} and \eqref{app:3.3} in \eqref{app:3.5}, after some algebraic manipulations, one obtains equation \eqref{app:3.7} reported at the top of the next page,
\begin{figure*}
	\begin{align}
		\label{app:3.7}
		\rho_3 &= \frac{1}{2} \begin{bmatrix}
			\Gamma \tilde{\rho}_e + \Lambda \tilde{\rho}_e \chi 
				& \Gamma \tilde{\rho}_e - \Lambda \tilde{\rho}_e \chi \\
	 		& & \\
			(I\otimes Z)\Gamma (I\otimes Z) \tilde{\rho}_e + (I\otimes Z)\Lambda (I\otimes Z) \tilde{\rho}_e \chi 
				& (I\otimes Z)\Gamma (I\otimes Z) \tilde{\rho}_e - (I\otimes Z)\Lambda (I\otimes Z) \tilde{\rho}_e \chi \\
	\end{bmatrix}.
\end{align}
\end{figure*}, with $\Gamma \in \mathbb{C}^{4\times4}$ and $\Lambda \in \mathbb{C}^{4\times4}$ defines as:
\begin{equation}
	\label{app:3.8}
	\Gamma= \begin{bmatrix}
			\rho_{\psi}^{11} I_{2\times2} & \rho_{\psi}^{21} I_{2\times2} \\ 
			\rho_{\psi}^{21} I_{2\times2} & \rho_{\psi}^{11} I_{2\times2} 
		\end{bmatrix},
\end{equation}
\begin{equation}
	\label{app:3.9}
		\Lambda= \begin{bmatrix}
	 			\rho_{\psi}^{12} I_{2\times2} & \rho_{\psi}^{22} I_{2\times2} \\ 
				\rho_{\psi}^{22} I_{2\times2} & \rho_{\psi}^{12} I_{2\times2}
			\end{bmatrix}.
\end{equation}

Finally, as shown in Fig.~\ref{Fig:A}, Alice jointly measures the pair of quantum states at her side, with $25\%$ chance of finding each of the four combinations $00,01,10,11$. Alice's measurement operation instantaneously fixes Bob's quantum state -- regardless of the distance between Alice and Bob -- as a consequence of the entanglement. However, Bob can only recover the original state after he correctly receives the pair of classical bits conveying the specific results of Alice's measurement. This further step projects $\rho_3$ on the subspaces described by the operators $\Pi_{ij} \in \mathbb{R}^{8\times8}=\ket{ij}\bra{ij} \otimes I_{2\times2}$, with $i,j \in \{0,1\}$.\\
More in detail, let us suppose that the measurement outcome is the one corresponding to the state $\ket{00}$. After the measurement, the global quantum state collapse into the state:
\begin{align}
	\label{app:3.10}
	\rho_4^{00} = \frac{\Pi_{00} \rho_3 \Pi_{00}^\dag}{\text{Tr}[{\Pi_{00}\rho_3 \Pi_{00}^\dag}]}.
\end{align}
As a consequence of its definition, $\Pi_{00}$ is equal to:
\begin{align}
\label{app:3.11}
	\Pi_{00}&=
	\begin{bmatrix}
		\overbrace{\begin{bmatrix} I_{2\times2} & 0_{2\times2} \\ 
		0_{2\times2} & 0_{2\times2}\end{bmatrix}}^{\ket{00}\bra{00}+\ket{01}\bra{01}} & 0_{4\times4}\\
		0_{4\times4} & o_{4\times4}
	\end{bmatrix}
\end{align}
By substituting \eqref{app:3.11} and \eqref{app:3.7} in \eqref{app:3.10}, and by exploiting the expressions of $\Gamma$ and $\Lambda$ given in \eqref{app:3.8} and \eqref{app:3.9}, after some algebraic manipulations, it can be recognized that \eqref{app:3.10} is equivalent to:
\begin{align}
	\label{app:3.12}
	\rho_4^{00}& =
		2 \ket{00}\bra{00} \otimes \left( \rho_{\psi}^{11} \tilde{\rho}_{e_{11}}+ \rho_{\psi}^{12} \tilde{\rho}_{e_{12}}+ \rho_{\psi}^{21} \tilde{\rho}_{e_{21}}+ \rho_{\psi}^{22} \tilde{\rho}_{e_{22}}\right),
\end{align}
where we utilized the block-structure of the matrix $\tilde{\rho}_e$ in terms of the $2\times 2$ sub-blocks $\{ \tilde{\rho}_{e_{ij}}\}_{i,j = 1,2}$.\\
From \eqref{app:3.12} -- by tracing out the composite Alice's state and by recalling that $\text{Tr}_{\mathcal{C}}(\mathcal{C}\otimes \mathcal{D})=\mathcal{D} \text{Tr}(\mathcal{C})$ -- it results that the density matrix $\rho_t$ of the teleported qubit at Bob's side, when the measurement outcome at Alice is equal to $\ket{00}$, is given by:
\begin{align}
	\label{app:3.13}
	\rho_t =2 \left( \rho_{\psi}^{11} \tilde{\rho}_{e_{11}}+ \rho_{\psi}^{12} \tilde{\rho}_{e_{12}}+ \rho_{\psi}^{21} \tilde{\rho}_{e_{21}}+ \rho_{\psi}^{22} \tilde{\rho}_{e_{22}}\right).
\end{align}
Hence by accounting for Definition~\ref{def:5.2}, the proof follows.

With the same reasoning, the lemma can be proved for different outcomes of the measurement process at Alice's side. As instance, let us suppose that the measurement outcome is the state $\ket{10}$. By reasoning as above, it results:
\begin{align}
	\label{app:3.14}
	\rho_4^{10}& =\frac{\Pi_{10}\rho_3 \Pi_{10}^\dag}{\text{Tr}[{\Pi_{10}\rho_3 \Pi_{10}^\dag}]}=\\&\nonumber
	=2 \ket{10}\bra{10} \otimes \left( \rho_{\psi}^{11} \tilde{\rho}_{e_{11}}- \rho_{\psi}^{12} \tilde{\rho}_{e_{12}}- \rho_{\psi}^{21} \tilde{\rho}_{e_{21}}+ \rho_{\psi}^{22} \tilde{\rho}_{e_{22}}\right).
\end{align}
From \eqref{app:3.14} -- by tracing out the composite Alice's state -- one obtains that the density matrix $\rho_t$ of the teleported qubit at Bob's side, after having applied the Z gate -- is given by:
\begin{align}
	\label{eq:5.18}
	\rho_t =2 Z\left( \rho_{\psi}^{11} \tilde{\rho}_{e_{11}}- \rho_{\psi}^{12} \tilde{\rho}_{e_{12}}- \rho_{\psi}^{21} \tilde{\rho}_{e_{21}}+ \rho_{\psi}^{22} \tilde{\rho}_{e_{22}}\right)Z.
\end{align}

\section{Proof of Proposition~\ref{prop:5.1}}
\label{app:4}
The average fidelity $\overline{F}^{\text{QS}}$ of the teleported qubit at Bob's side can be evaluated by averaging the conditional fidelity $F^{\text{QS}}(\theta, \phi) \eqdef \bra{\psi}\rho_t \ket{\psi}$ on all the possible values of the qubit $\ket{\psi}$, i.e.:
\begin{align}
	\label{app:4.1}
	\overline{F}^{\text{QS}} &= \frac{1}{4\pi}\int_{0}^\pi d\theta \int_{0}^{2\pi} F^{\text{QS}}(\theta, \phi) \sin(\theta) d\phi = \nonumber \\
	&= \frac{1}{4\pi}\int_{0}^\pi d\theta \int_{0}^{2\pi} \bra{\psi}\rho_t \ket{\psi}\sin(\theta) d\theta d\phi
\end{align}
By adopting a quantum switch for the entanglement distribution scheme, from Corollary~\ref{cor:4.1} the density matrix of the entangled pair $\rho_e^{\text{QS}}$ at the output of the quantum switch coincides with $\rho_e$ whenever the measurement of the control qubit $\ket{\varphi_c}$ provides as outcome the one corresponding to the state $\ket{-}$. And this outcome is obtained with probability $pq$. As a consequence, by substituting $\rho_e^{\text{QS}} = \rho_e$ in \eqref{eq:5.3} of Lemma~\ref{lem:5.1}, it results $\rho_t=\rho_{\psi}$. Hence, from \eqref{app:4.1}, the average fidelity $\overline{F}_{\ket{-}}^{\text{QS}}$ given that the control qubit $\ket{\varphi_c}$ is measured into state $\ket{-}$ is equal to $1$.

Conversely, whenever the measurement of the control qubit $\ket{\varphi_c}$ provides as outcome the one corresponding to the state $\ket{+}$, from Corollary~\ref{cor:4.1} the density matrix of the entangled pair $\rho_e^{\text{QS}}$ at the output of the quantum switch is given by \eqref{eq:4.5}. And this outcome is obtained with probability $(1-pq)$. As a consequence, by supposing without any loss of generality that $\mathbf{1}_{00}^A=1$\footnote{Whenever the indicator function of the measurement process at Alice is different from $\mathbf{1}_{00}^A=1$, all the above analysis continues to hold, since it is sufficient to single out the corresponding value of $\rho_t$ in \eqref{eq:5.3}.}, and by substituting the expression \eqref{eq:4.5} of $\rho_e^{\text{QS}}$ in \eqref{eq:5.3}, after some algebraic manipulations it results:
\begin{align}
	\label{app:4.2}
	\overline{F}_{\ket{+}}^{\text{QS}} & =\frac{1}{4\pi}\int_{0}^\pi d\theta \int_{0}^{2\pi} \bra{\psi}\rho_t \ket{\psi}\sin(\theta) d\theta d\phi= \nonumber \\
		& = \frac{1}{4\pi (1-pq)}\int_{0}^\pi d\theta \int_{0}^{2\pi} \left[(1-p)-q(1-p)\sin^{2}(\theta)+\right. \nonumber \\
		& \quad \left. + p(1-q)\sin^{2}(\theta)\cos^{2}(\phi)\right]\sin(\theta) d\theta d\phi.
\end{align}
and the proof follows by solving \eqref{app:4.2}.

\section{Proof of Corollary~\ref{cor:5.1}}
\label{app:5}

According to the result of Corollary~\ref{cor:4.2}, the density matrix of the entangled pair $\rho_e^{\text{CT}}$ when no quantum switch is adopted is given by \eqref{eq:4.6}.
As a consequence, by assuming without any loss of generality that $\mathbf{1}_{00}^A=1$\footnote{When the indicator function of the measurement process at Alice is different from $\mathbf{1}_{00}^A=1$, all the above analysis continues to hold, since it is sufficient to single out the corresponding value of $\rho_t$ in \eqref{eq:5.3}.}, and by substituting $\rho_e^{\text{CT}}$ in \eqref{eq:5.3} of Lemma~\ref{lem:5.1}, it results that the average Fidelity $\overline{F}^{\text{CT}}$ when no quantum switch is adopted is given by:
\begin{align}
\nonumber
\overline{F}^{\text{CT}} &
=\frac{1}{4\pi}\int_{0}^\pi d\theta \int_{0}^{2\pi} F(\theta, \phi)\sin(\theta) d\theta d\phi=\\&\nonumber
= \frac{1}{4\pi (1-pq)}\int_{0}^\pi d\theta \int_{0}^{2\pi} \left[(1-p)-q(1-2p)\sin^{2}(\theta)+\right. \\&\nonumber \left. + p(1-2q)\sin^{2}(\theta)\cos^{2}(\phi)\right]\sin(\theta) d\theta d\phi=
\\&=
\frac{3-2p-2q+2pq}{3}.
\label{app:5.1}
\end{align}

The proof easily follows by considering that \eqref{eq:5.5} can be equivalently written in a compact form as:
\begin{equation}
\label{app:5.2}
\overline{F}^{\text{QS}}= pq \overline{F}^{\text{QS}}_{\ket{-}}+ (1-pq)\overline{F}^{\text{QS}}_{ \ket{+}}=\frac{3-2p-2q+4pq}{3}.
\end{equation} 
In fact, by comparing \eqref{app:5.2} with \eqref{app:5.1}, one obtains that for every $p, q \neq 0$, $\overline{F}^{\text{QS}}>\overline{F}^{\text{CT}}$.

\small
\bibliographystyle{IEEEtran}
\bibliography{quantum-07}

\begin{thebibliography}{10}
\providecommand{\url}[1]{#1}
\csname url@samestyle\endcsname
\providecommand{\newblock}{\relax}
\providecommand{\bibinfo}[2]{#2}
\providecommand{\BIBentrySTDinterwordspacing}{\spaceskip=0pt\relax}
\providecommand{\BIBentryALTinterwordstretchfactor}{4}
\providecommand{\BIBentryALTinterwordspacing}{\spaceskip=\fontdimen2\font plus
\BIBentryALTinterwordstretchfactor\fontdimen3\font minus
  \fontdimen4\font\relax}
\providecommand{\BIBforeignlanguage}[2]{{%
\expandafter\ifx\csname l@#1\endcsname\relax
\typeout{** WARNING: IEEEtran.bst: No hyphenation pattern has been}%
\typeout{** loaded for the language `#1'. Using the pattern for}%
\typeout{** the default language instead.}%
\else
\language=\csname l@#1\endcsname
\fi
#2}}
\providecommand{\BIBdecl}{\relax}
\BIBdecl

\bibitem{SalEblChi-18}
S.~{Salek}, D.~{Ebler}, and G.~{Chiribella}, ``{Quantum communication in a
  superposition of causal orders},'' \emph{arXiv e-prints}, p.
  arXiv:1809.06655, Sep. 2018.

\bibitem{ChiKri-19}
G.~Chiribella and H.~Kristjánsson, ``Quantum shannon theory with superpositions
  of trajectories,'' \emph{Proceedings of the Royal Society A}, vol. 475, 2019.

\bibitem{AbbWecHor-18}
A.~A. {Abbott}, J.~{Wechs}, D.~{Horsman} \emph{et~al.}, ``{Communication
  through coherent control of quantum channels},'' \emph{arXiv e-prints}, p.
  arXiv:1810.09826, Oct. 2018.

\bibitem{ChiDarPer-13}
G.~Chiribella, G.~M. D'Ariano, P.~Perinotti, and B.~Valiron, ``Quantum
  computations without definite causal structure,'' \emph{Phys. Rev. A},
  vol.~88, p. 022318, Aug. 2013.

\bibitem{ColDarFac-12}
T.~Colnaghi, G.~M. D'Ariano, S.~Facchini, and P.~Perinotti, ``Quantum
  computation with programmable connections between gates,'' \emph{Physics
  Letters A}, vol. 376, no.~45, pp. 2940 -- 2943, 2012.

\bibitem{AraCosBru-14}
M.~Ara\'ujo, F.~Costa, and {\v C}.~Brukner, ``{Computational Advantage from
  Quantum-Controlled Ordering of Gates},'' \emph{Phys. Rev. Lett.}, vol. 113,
  p. 250402, Dec. 2014.

\bibitem{Chi-12}
G.~Chiribella, ``Perfect discrimination of no-signalling channels via quantum
  superposition of causal structures,'' \emph{Phys. Rev. A}, vol.~86, p.
  040301, Oct. 2012.

\bibitem{WakSoeMur-19}
E.~Wakakuwa, A.~Soeda, and M.~Murao, ``{Complexity of Causal Order Structure in
  Distributed Quantum Information Processing: More Rounds of Classical
  Communication Reduce Entanglement Cost},'' \emph{Phys. Rev. Lett.}, vol. 122,
  p. 190502, May 2019.

\bibitem{OreCosBru-12}
O.~Oreshkov, F.~Costa, and {\v C}.~Brukner, ``Quantum correlations with no
  causal order,'' \emph{Nature Communications}, vol.~3, pp. 1092 EP --, Oct.
  2012.

\bibitem{FeiAraBru-15}
A.~Feix, M.~Ara\'ujo, and {\v C}.~Brukner, ``Quantum superposition of the order
  of parties as a communication resource,'' \emph{Phys. Rev. A}, vol.~92, p.
  052326, Nov. 2015.

\bibitem{GueFeiAra-16}
P.~A. Gu\'erin, A.~Feix, M.~Ara\'ujo, and {\v C}.~Brukner, ``{Exponential
  Communication Complexity Advantage from Quantum Superposition of the
  Direction of Communication},'' \emph{Phys. Rev. Lett.}, vol. 117, p. 100502,
  Sep 2016.

\bibitem{ProMoqAra-15}
L.~M. Procopio, A.~Moqanaki, M.~Ara{\'u}jo \emph{et~al.}, ``Experimental
  superposition of orders of quantum gates,'' \emph{Nature Communications},
  vol.~6, pp. 7913 EP --, Aug. 2015.

\bibitem{RubRozFei-17}
G.~Rubino, L.~A. Rozema, A.~Feix \emph{et~al.}, ``Experimental verification of
  an indefinite causal order,'' \emph{Science Advances}, vol.~3, no.~3, 2017.

\bibitem{RubTozMas-19}
G.~Rubino, L.~A. Rozema, F.~Massa \emph{et~al.}, ``{Experimental Entanglement
  of Temporal Orders},'' in \emph{Quantum Information and Measurement (QIM) V:
  Quantum Technologies}, 2019.

\bibitem{EblSalChi-18}
D.~Ebler, S.~Salek, and G.~Chiribella, ``Enhanced communication with the
  assistance of indefinite causal order,'' \emph{Phys. Rev. Lett.}, vol. 120,
  p. 120502, Mar. 2018.

\bibitem{GueRubBru-19}
P.~A. Gu\'erin, G.~Rubino, and {\v C}.~Brukner, ``Communication through
  quantum-controlled noise,'' \emph{Phys. Rev. A}, vol.~99, p. 062317, June
  2019.

\bibitem{GosRomWhi-18}
K.~{Goswami}, J.~{Romero}, and A.~G. {White}, ``{Communicating via
  ignorance},'' \emph{arXiv e-prints}, p. arXiv:1807.07383, Jul. 2018.

\bibitem{WeiNorZha-19}
K.~Wei, N.~Tischler, S.-R. Zhao \emph{et~al.}, ``{Experimental Quantum
  Switching for Exponentially Superior Quantum Communication Complexity},''
  \emph{Phys. Rev. Lett.}, vol. 122, p. 120504, Mar. 2019.

\bibitem{ChiBanSan-18}
G.~{Chiribella}, M.~{Banik}, S.~{Sankar Bhattacharya} \emph{et~al.},
  ``{Indefinite causal order enables perfect quantum communication with zero
  capacity channel},'' \emph{arXiv e-prints}, p. arXiv:1810.10457, Oct. 2018.

\bibitem{GosGiaKew-18}
K.~Goswami, C.~Giarmatzi, M.~Kewming \emph{et~al.}, ``{Indefinite Causal Order
  in a Quantum Switch},'' \emph{Phys. Rev. Lett.}, vol. 121, p. 090503, Aug.
  2018.

\bibitem{GuoHuHou-18}
Y.~{Guo}, X.-M. {Hu}, Z.-B. {Hou} \emph{et~al.}, ``{Experimental investigating
  communication in a superposition of causal orders},'' \emph{arXiv e-prints},
  p. arXiv:1811.07526, Nov. 2018.

\bibitem{CacCalVanHan-19}
A.~S. Cacciapuoti, M.~Caleffi, R.~Van~Meter, and L.~Hanzo, ``When entanglement
  meets classical communications: Quantum teleportation for the quantum
  internet (invited paper),'' \emph{ArXiv e-prints}, p. arXiv:1907.06197, July
  2019.

\bibitem{CalCacBia-18}
M.~Caleffi, A.~S. Cacciapuoti, and G.~Bianchi, ``{Quantum Internet: From
  Communication to Distributed Computing!}'' in \emph{NANOCOM '18}, 2018, pp.
  1--4.

\bibitem{CacCalTaf-18}
A.~S. Cacciapuoti, M.~Caleffi, F.~Tafuri, F.~S. Cataliotti, S.~Gherardini, and
  G.~Bianchi, ``{Quantum Internet: Networking Challenges in Distributed Quantum
  Computing},'' \emph{ArXiv e-prints}, p. arXiv:1810.08421, Oct. 2018.

\bibitem{BenWie-92}
C.~H. Bennett and S.~J. Wiesner, ``{Communication via one- and two-particle
  operators on Einstein-Podolsky-Rosen states},'' \emph{Physical Review
  Letters}, vol.~69, pp. 2881--2884, Nov. 1992.

\bibitem{BenBraCre-93}
C.~H. Bennett, G.~Brassard, C.~Cr\'epeau \emph{et~al.}, ``{Teleporting an
  unknown quantum state via dual classical and Einstein-Podolsky-Rosen
  channels},'' \emph{Physical Review Letters}, vol.~70, pp. 1895--1899, Mar.
  1993.

\bibitem{CacCal-19}
A.~S. Cacciapuoti and M.~Caleffi, ``{Toward the Quantum Internet: A
  Directional-dependent Noise Model for Quantum Signal Processing (Invited
  Paper)},'' in \emph{ICASSP'19}, 2019.

\bibitem{Wil-13}
M.~M. Wilde, \emph{{Quantum Information Theory}}, 1st~ed.\hskip 1em plus 0.5em
  minus 0.4em\relax New York, NY, USA: Cambridge University Press, 2013.

\bibitem{EinPodRos-35}
A.~Einstein, B.~Podolsky, and N.~Rosen, ``{Can Quantum-Mechanical Description
  of Physical Reality Be Considered Complete?}'' \emph{Physical Review},
  vol.~47, pp. 777--780, May 1935.

\bibitem{KwiMatWei-95}
P.~G. Kwiat, K.~Mattle, H.~Weinfurter, A.~Zeilinger, A.~V. Sergienko, and
  Y.~Shih, ``{New High-Intensity Source of Polarization-Entangled Photon
  Pairs},'' \emph{Phys. Rev. Lett.}, vol.~75, pp. 4337--4341, Dec. 1995.

\bibitem{NorBla-17}
T.~E. Northup and R.~Blatt, ``Quantum information transfer using photons,''
  \emph{Nature Photonics}, vol.~8, pp. 356 EP --, Apr. 2014.

\bibitem{MukPat-19}
C.~{Mukhopadhyay} and A.~K. {Pati}, ``{Superposition of causal order enables
  perfect quantum teleportation with very noisy singlets},'' \emph{arXiv
  e-prints}, p. arXiv:1901.07626, Jan. 2019.

\bibitem{NieChu-11}
M.~A. Nielsen and I.~L. Chuang, \emph{Quantum Computation and Quantum
  Information}, {10th Anniversary}~ed.\hskip 1em plus 0.5em minus 0.4em\relax
  Cambridge University Press, 2011.

\bibitem{Joz-94}
R.~Jozsa, ``{Fidelity for Mixed Quantum States},'' \emph{Journal of Modern
  Optics}, vol.~41, pp. 2315--2323, 1994.

\bibitem{VanMet-14}
R.~Van{ }Meter, \emph{Quantum Networking}.\hskip 1em plus 0.5em minus
  0.4em\relax Wiley-ISTE, Apr. 2014.

\bibitem{MasPop-95}
S.~Massar and S.~Popescu, ``{Optimal Extraction of Information from Finite
  Quantum Ensembles},'' \emph{Phys. Rev. Lett.}, vol.~74, pp. 1259--1263, Feb.
  1995.

\bibitem{Ben-96}
C.~H. Bennett, G.~Brassard, S.~Popescu \emph{et~al.}, ``{Purification of Noisy
  Entanglement and Faithful Teleportation via Noisy Channels},'' \emph{Phys.
  Rev. Lett.}, vol.~76, pp. 722--725, Jan. 1996.

\bibitem{Han-QERC-18}
Z.~{Babar}, D.~{Chandra}, H.~V. {Nguyen}, P.~{Botsinis}, D.~{Alanis}, S.~X.
  {Ng}, and L.~{Hanzo}, ``{Duality of Quantum and Classical Error Correction
  Codes: Design Principles and Examples},'' \emph{IEEE Communications Surveys
  Tutorials}, vol.~21, no.~1, pp. 970--1010, 2019.

\end{thebibliography}

\begin{IEEEbiography}
[{\includegraphics[width=1in,height=1.25in,clip,keepaspectratio]{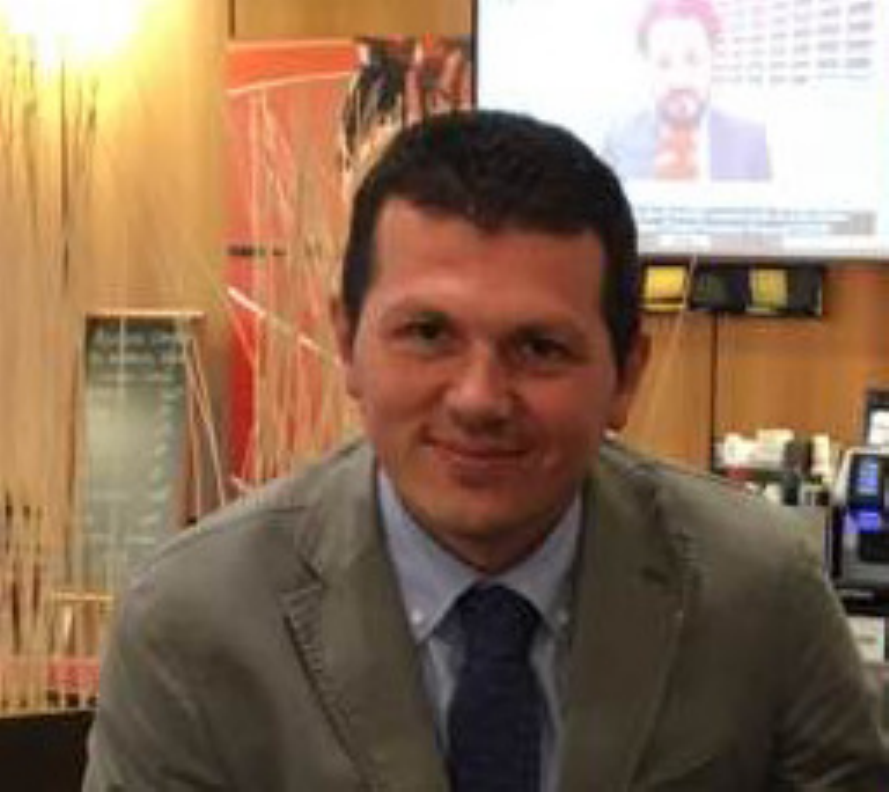}}]{Marcello Caleffi} (M'12, SM'16) received the M.S. degree (summa cum laude) in computer science engineering from the University of Lecce, Lecce, Italy, in 2005, and the Ph.D. degree in electronic and telecommunications engineering from the University of Naples Federico II, Naples, Italy, in 2009. Currently, he is with the DIETI Department, University of Naples Federico II, and with the National Laboratory of Multimedia Communications, National Inter-University Consortium for Telecommunications (CNIT). From 2010 to 2011, he was with the Broadband Wireless Networking Laboratory at Georgia Institute of Technology, Atlanta, as visiting researcher. In 2011, he was also with the NaNoNetworking Center in Catalunya (N3Cat) at the Universitat Politecnica de Catalunya (UPC), Barcelona, as visiting researcher. Since July 2018, he held the Italian national habilitation as \textit{Full Professor} in Telecommunications Engineering. His work appeared in several premier IEEE Transactions and Journals, and he received multiple awards, including \textit{best strategy} award, \textit{most downloaded article} awards and \textit{most cited article} awards. Currently, he serves as associate technical editor for IEEE Communications Magazine and as editor for IEEE Communications Letters and. He has served as Chair, TPC Chair, Session Chair, and TPC Member for several premier IEEE conferences. In 2016, he was elevated to IEEE Senior Member and in 2017 he has been appointed as Distinguished Lecturer from the \textit{IEEE Computer Society}. In December 2017, he has been elected Treasurer of the Joint \textit{IEEE VT/ComSoc Chapter Italy Section}. In December 2018, he has been appointed member of the IEEE \textit{New Initiatives Committee}.
\end{IEEEbiography}

\begin{IEEEbiography}
[{\includegraphics[width=1in,height=1.25in,clip,keepaspectratio]{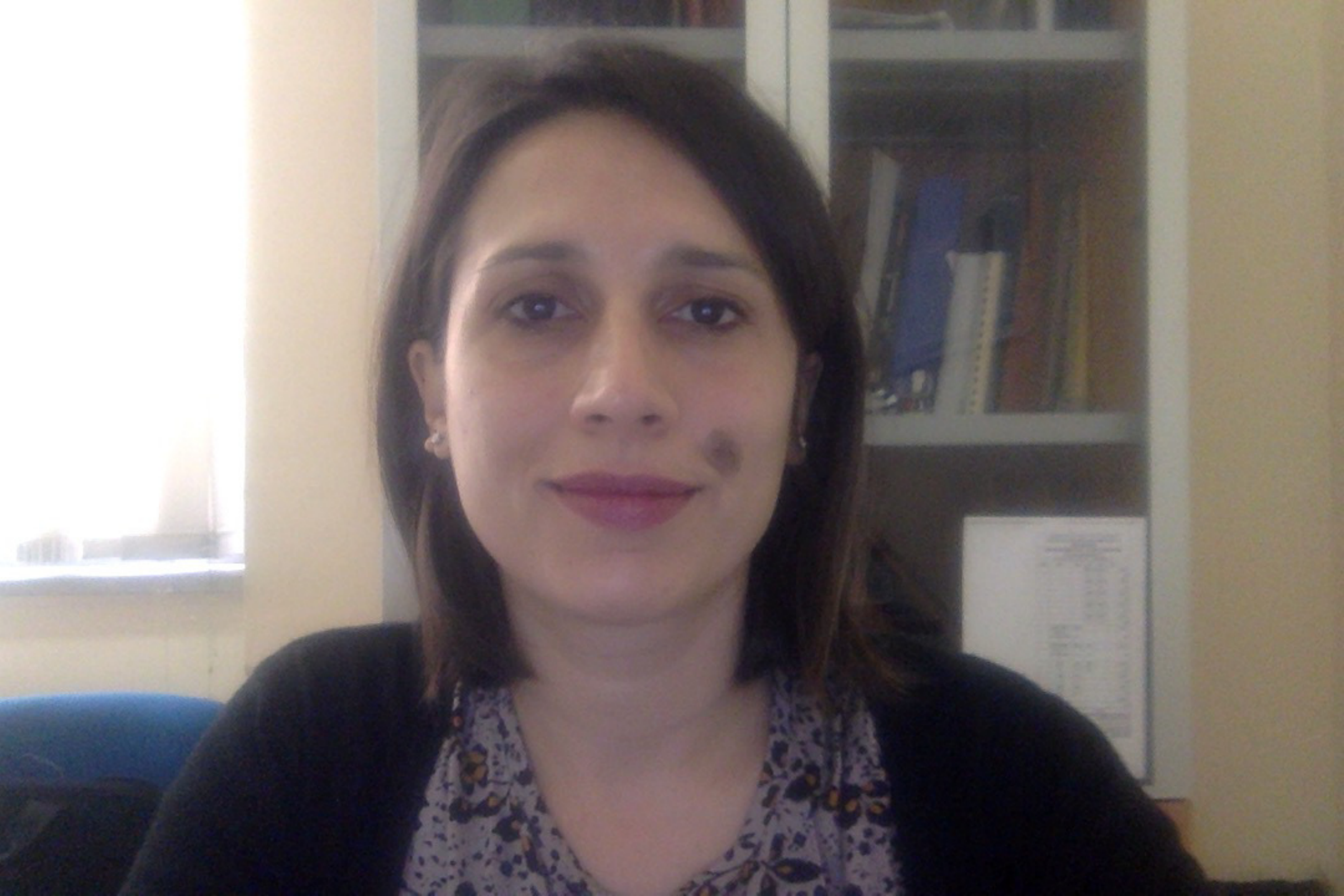}}]{Angela Sara Cacciapuoti} (M'10, SM'16) received the Ph.D. degree in Electronic and Telecommunications Engineering in 2009, and the \textit{Laurea summa cum laude} in Telecommunications Engineering in 2005, both from the University of Naples Federico II. Since April 2017, she held the Italian Habilitation as ``Associate Professor'' in Telecommunications Engineering and since July 2018, she held the Italian Habilitation as ``Full Professor'' in Telecommunications Engineering. Currently, she is a Tenure-Track Assistant Professor at the Department of Electrical Engineering and Information Technology, University of Naples Federico II. Prior to that, she was a visiting researcher at the Georgia Institute of Technology (USA) and at the NaNoNetworking Center in Catalunya (N3Cat), School of Electrical Engineering, Universitat Politecnica de Catalunya (Spain). Her work appeared in the first tier IEEE journals and she received different awards including the elevation to the grade of IEEE Senior Member in February 2016, most downloaded article and most cited article awards, and outstanding young faculty/researcher fellowships for conducting research abroad. Angela Sara serves as Editor/Associate Editor for the journals: IEEE Trans. on Communications, IEEE Communications Letters, Computer Networks (Elsevier) Journal and IEEE Access. She was awarded with the 2017 Exemplary Editor Award of IEEE Communications Letters. Since 2016, she is a board member of the IEEE ComSoc Young Professionals (YPs) Standing Committee and, since 2018 of the IEEE ComSoc Women in Communications Engineering (WICE) Standing Committee. In February 2017, she has been appointed Award Co-Chair of the N2Women Board and in July 2017, she has been elected Treasurer of the IEEE Women in Engineering (WIE) Affinity Group (AG) of the IEEE Italy Section.
\end{IEEEbiography}

\balance

\end{document}